\documentclass{llncs}
\usepackage{babel}

\usepackage{graphicx}

\usepackage[T1]{fontenc}
\usepackage{cite}
\usepackage{color}
\usepackage[colorlinks=true]{hyperref}

\usepackage{amsmath,amssymb,amsfonts}
\usepackage{amstext}
\usepackage{xspace}
\usepackage{microtype}
\usepackage{thmtools}
\usepackage{doi}
\usepackage{uppaal}

\usepackage{tikz}
\usetikzlibrary{automata, arrows.meta, positioning}
\usepackage[font=scriptsize,labelfont=bf]{caption}

\usepackage{braket}
\usepackage{nicefrac}

\usepackage{listings}
\usepackage[textsize=scriptsize,textwidth=4.3cm]{todonotes}

\lstdefinelanguage{aald}
	{morekeywords={system, implementation},
	sensitive=false,
	morecomment=[l]{//},
	morecomment=[s]{/*}{*/},
	morestring=[b]",
}

\definecolor{delim}{RGB}{20,105,176}
\definecolor{numb}{RGB}{106, 109, 32}
\definecolor{string}{rgb}{0.64,0.08,0.08}

\lstdefinelanguage{json}{
    numbers=left,
    numberstyle=\small,
    frame=single,
    rulecolor=\color{black},
    showspaces=false,
    showtabs=false,
    breaklines=true,
    postbreak=\raisebox{0ex}[0ex][0ex]{\ensuremath{\color{gray}\hookrightarrow\space}},
    breakatwhitespace=true,
    basicstyle=\ttfamily\small,
    upquote=true,
    morestring=[b]",
    stringstyle=\color{string},
    tabsize=2,
    literate=
     *{0}{{{\color{numb}0}}}{1}
      {1}{{{\color{numb}1}}}{1}
      {2}{{{\color{numb}2}}}{1}
      {3}{{{\color{numb}3}}}{1}
      {4}{{{\color{numb}4}}}{1}
      {5}{{{\color{numb}5}}}{1}
      {6}{{{\color{numb}6}}}{1}
      {7}{{{\color{numb}7}}}{1}
      {8}{{{\color{numb}8}}}{1}
      {9}{{{\color{numb}9}}}{1}
      {\{}{{{\color{delim}{\{}}}}{1}
      {\}}{{{\color{delim}{\}}}}}{1}
      {[}{{{\color{delim}{[}}}}{1}
      {]}{{{\color{delim}{]}}}}{1},
}

\usepackage[linesnumbered,ruled,vlined]{algorithm2e}
\usepackage{makecell}
\usepackage{enumitem}
\definecolorseries{test}{rgb}{grad}[rgb]{.95,.55,.55}{11,11,17}
\resetcolorseries[10]{test}
\newcommand{\addtodoeditor}[1]{%
    \colorlet{#1}{test!!+!50}
    \expandafter\newcommand\csname#1\endcsname [1]{%
        \todo[color=#1,size=\tiny,inline]{\sffamily\textbf{\uppercase{#1}:}
    ##1}\xspace%
    }
    \expandafter\newcommand\csname#1i\endcsname [1]{%
        \todo[inline, color=#1]{\sffamily\textbf{\uppercase{#1}:} ##1}\xspace%
    }
}

\addtodoeditor{mo}
\addtodoeditor{mm}
\addtodoeditor{mz}
\usepackage[group-separator={,}]{siunitx}

\newcommand{\card}[1]{\mathit{|#1|}}

\newcommand{\partition}{\mathit{prt}}

\newcommand{\task}[1]{\textsf{#1}}

\newcommand{\affinity}{\mathit{aff}}

\newcommand{\ie}{\mbox{i.e.,\@}\xspace}%
\newcommand{\eg}{\mbox{e.g.,\@}\xspace}%
\newcommand{\wrt}{\mbox{w.r.t.\@}\xspace}%
\newcommand{\nonl}{\renewcommand{\nl}{\let\nl\oldnl}}% 
\newcommand{\isep}{\mathrel{{.}\,{.}}\nobreak}
\begin{document}
\setlength{\abovedisplayskip}{1pt}
\setlength{\belowdisplayskip}{1pt}
\date{}

\makeatletter
\def\ps@pprintTitle{%
  \let\@oddhead\@empty
  \let\@evenhead\@empty
  \let\@oddfoot\@empty
  \let\@evenfoot\@oddfoot
}
\makeatother

\title{Scalable Computation of Inter-Core Bounds Through Exact Abstractions}

\author{Mohammed Aristide Foughali\inst{1,3}%\orcidID{0000-1111-2222-3333} 
\and
Marius Mikučionis\inst{2}
\and
Maryline Zhang\inst{1}
}
%
%\authorrunning{F. Author et al.}
% First names are abbreviated in the running head.
% If there are more than two authors, 'et al.' is used.
%
\institute{Université Paris Cité, CNRS, IRIF, F-75013, Paris, France\\
\and
Department of Computer Science, Aalborg University, Denmark
\and
Email \email{foughali@irif.fr}}
\maketitle              % typeset the header of the contribution

\begin{abstract}
Real-time systems (RTSs) are at the heart of numerous safety-critical applications. %
An RTS typically consists of a set of real-time tasks (the software) that execute on a multicore shared-memory 
platform (the hardware) following a \emph{scheduling policy}. %
In an RTS, computing \emph{inter-core bounds}, \ie bounds separating events produced by tasks on different cores,   
is crucial. %
While efficient techniques to over-approximate such bounds exist, little has been proposed to compute their exact values. % 
Given an RTS with a set of cores $C$ and a set of tasks $T$, 
under partitioned fixed-priority scheduling with limited preemption, 
a recent work by Foughali, Hladik and Zuepke (FHZ) models tasks with affinity $c$ (\ie allocated to core $c\in C$) as a \uppaal timed automata (TA) network $N_c$. %
For each core $c$ in $C$, $N_c$ integrates blocking (due to data sharing) using tight analytical formulae. %
Through compositional model checking, FHZ achieved a substantial gain in scalability for bounds \emph{local to a core}. %
However, computing inter-core bounds for some events of interest $E$, produced by a subset of tasks $T_E\subseteq T$ with different affinities $C_{E}\subseteq C$, requires model checking $\mathit{N_{E} = ||_{c\in C_{E}} N_c}$, \ie the parallel composition of all TA networks $N_c$ for each $c\in C_{E}$, which produces a large, often intractable, state space. %
In this paper, we present a new scalable approach based on \emph{exact abstractions} to compute exact inter-core bounds in a \emph{schedulable} RTS, under the assumption that tasks in $T_E$ have distinct affinities, \ie $\card{T_E} = \card{C_E}$. %
We develop a novel algorithm, leveraging a new query that we implement in \uppaal, that computes for each TA network $N_c$ in $\mathit{N_{E}}$ an abstraction $\mathcal{A}(N_c)$ preserving the \emph{exact} intervals within which events occur on $c$. %
Then, we model check $\mathit{\mathcal{A}(N_{E}) = ||_{c\in C_{E}} \mathcal{A}(N_c)}$ (instead of $\mathit{N_{E}}$), therefore drastically reducing the state space. %
The scalability of our approach is demonstrated on the WATERS 2017 industrial challenge, %
for which we efficiently compute various types of inter-core bounds where FHZ fails to scale. %

\keywords{Timed automata  \and Model checking \and Real-time systems}
\end{abstract}

\section{Introduction}\label{intro} 
Real-time systems (RTSs) underly many safety-critical applications, spanning areas such as robotics and automotive industry. %
An RTS typically boils down to a set of complex real-time \emph{tasks} (the software) that execute on a \emph{multicore shared-memory} platform (the hardware). %
Tasks, that   
must obey stringent timing constraints (\emph{deadlines}) under limited resources, \ie a small number of cores, and concurrency protocols, execute following a scheduling policy. %
Verifying that an RTS meets the \emph{schedulability} requirement is crucial. %
An RTS is schedulable, under a given scheduling policy, if each of its tasks is schedulable, \ie it always finishes its execution before its deadline~\cite{buttazzo2011hard}. %
Response-time analysis (RTA)~\cite{davis:2008,DBLP:conf/ecrts/MaidaBB22,DBLP:journals/tc/BiniB04,nasri2017exact} is a popular approach to verify schedulability, through computing the worst-case response time (WCRT) of each task. %

Schedulability is, however, not all what an RTS is about. %
In a schedulable RTS, one still needs to compute bounds separating some events of interest. %
To put this in the most generic way, let $T_E$ be the set of tasks producing some events in the set $E$. %
The goal is to compute, in a scalable way, exact bounds separating the production of events in $E$ following some requirement. %
For example, $E = \{e_1, e_2, e_3, e_4\}$ and we want to quantify precisely the minimum and maximum amount of time, in all possible scenarios, between each production of $e_1$ and each production of $e_4$, with $e_2$ and $e_3$ occurring concurrently in between. %
Solving this problem is hard, in particular if tasks in $T_E$ execute on different cores (\ie computing \emph{inter-core} bounds). %
Indeed, RTA-based techniques, specific to schedulability, are not suitable for inter-core bounds. %
In contrast, model checking is a natural candidate but suffers from state-space explosion. %
To illustrate this, consider a recent work by Foughali, Hladik and Zuepke~\cite{DBLP:journals/jsa/FoughaliHZ23} (FHZ hereafter). %
Given an RTS with \emph{periodic} tasks and \emph{partitioned fixed-priority} (P-FP) scheduling with \emph{limited preemption} (Sect.~\ref{prel}), the behavior of tasks with affinity $c$ (\ie allocated to core $c$), is modeled    
as $\mathit{N_c}$, a \uppaal timed automata (TA) network~\cite{larsen1997uppaal}. %
Since $N_c$ analytically integrates \emph{blocking} due to data sharing, WCRTs of tasks with affinity $c$ are computed on the state space of $N_c$ only in an efficient manner (\eg more than seven times faster than the Schedule Abstract Graph~\cite{nasri2017exact} on the WATERS 2017 case study~\cite{hamann2017waters}). % 
However, for an inter-core bound, 
 FHZ suggest to compute it 
on the parallel composition $\mathit{N_{E} = ||_{c\in C_{E}} N_c}$, where $C_{E}$ is the smallest subset of cores covering the affinities of tasks in $T_E$. %
Though this remains compositional, building the state space of $\mathit{N_E}$ does not scale in practice (Sect.~\ref{eval}). Many works focus on \emph{end-to-end latencies}~\cite{feiertag2009compositional,martinez2020end,gunzel2023compositional}, a special case of the problem above, %
for which efficient yet non-exact procedures are proposed (more in Sect.~\ref{rw}). % 

\textbf{Contributions.} We present in this paper a scalable approach to compute exact inter-core bounds %
in a schedulable RTS following the FHZ model, under the assumption that tasks in $T_E$ have distinct  affinities (\ie $\card{T_E} = \card{C_E}$).  %
We devise a novel algorithm, using a new query that we implement in \uppaal, that computes from every core network $N_c$, $\mathit{c\in C_{E}}$, an \emph{exact abstraction} $\mathcal{A}(N_c)$ that preserves the exact intervals within which events in $E$ occur on $c$. %
Accordingly, we compute exact inter-core bounds on the state space of $\mathit{\mathcal{A}(N_{E}) = ||_{c\in C_{E}} \mathcal{A}(N_c)}$, instead of $\mathit{N_{E} = ||_{c\in C_{E}} N_c}$. %
$\mathit{\mathcal{A}(N_{E})}$ underlies a drastically smaller state space compared to $\mathit{N_{E}}$, and therefore scalability is significantly improved. % 
Our approach is successfully evaluated on the WATERS 2017 industrial challenge~\cite{hamann2017waters}. %
 
\textbf{Outline.} %
In Sect.~\ref{prel}, we introduce models and tools used in this paper, namely TA, \uppaal and the FHZ model. %
Sect.~\ref{appr} presents the core of our contribution, \ie a novel algorithm to compute exact abstractions from the TA networks of FHZ. %
We demonstrate the scalability of our approach in Sect.~\ref{eval}. %
Finally, we compare with related work (Sect.~\ref{rw}) and wrap up with concluding remarks (Sect.~\ref{concl}).

\section{Preliminaries}\label{prel}
\subsection{\uppaal}\label{upp}

\uppaal~\cite{larsen1997uppaal} is a state-of-the-art model checker. %
It features a modeling language, based on an extension of timed automata (TA)~\cite{alur1994theory}, and a query (property) language, based on a subset of the Computation Tree Logic (CTL)~\cite{clarke1986automatic}. %
 
\subsubsection{Modeling Language}
\paragraph{TA syntax.} a timed automaton $\mathit{A}$ is a tuple $\langle\Sigma,L,\ell_0,X,E,I\rangle$, where:

\begin{description}[style=unboxed]
\item[actions] $\Sigma$ is a finite set of actions (including the \emph{silent} action $\epsilon$), 
\item[locations] $L$ is a finite set of locations,
\item[initial location] $\ell_0\in L$ is the initial location,
\item[clocks] $X$ a finite set of real-valued clocks, 
\item[edges] $E$ a set of edges $(\ell_i,g,a,u,\ell_j)$ between locations $\ell_i,\ell_j \in L$, with guard $g\in G$, action label $a\in\Sigma$ and $u\subseteq X$ reset expression over clocks, 
\item[invariants] $\mathit{I}$ a location invariant $\forall\ell\in L$: $\mathit{I}(\ell)\in G$. %
\end{description}

Constraints $G$ are conjunctions of the atomic form $\{x_i-x_j\prec c_{i,j}\;|\;0\leq i,j\leq \card{X}\}$ over clock variables $x_{i},x_{j}\in X$, with $x_0$ a special variable that is always equal to zero, $c_{i,j}\in\mathbb{Z}$ an integer bound and $\prec\in\{<,\leq\}$. % 
Constraints of the form $x_i - x_0 \leq c_{i,0}$ (resp. $x_0 - x_i \leq c_{0,i}$) are written simply as $x_i \leq c_{i,0}$ (resp. $x_i \geq -c_{0,i}$) for some $i\neq 0$. %
Similarly, a tautology $\top$ can be expressed as $x_0 - x_i \leq 0$ for some $i\in 0\isep \card{X}$. %
A location $\ell$ with an invariant equivalent to $\top$ is referred to as \emph{invariant free}. %
For simplicity, guards and invariants equivalent to $\top$ are not explicitly represented in this paper. %

\paragraph{TA semantics.} $A=\langle\Sigma,L,\ell_0, X,E,I\rangle$ is defined by a Timed Labeled Transition System $\mathit{TLTS}=\langle \Sigma,S,s_0,T\rangle$:
\begin{description}[style=unboxed]
\item[state] $\langle\ell,\overline{v}\rangle\in S$ consists of a location $\ell\in L$ and a valuation vector $\overline{v}$  assigning a non-negative real $\overline{v}[x]\in\mathbb{R}_{\geq 0}$ to each clock variable $x\in X$,
\item[initial state] $s_0=\langle\ell_0,\overline{0}\rangle\in S$ 
with $\overline{0}[x]=0$ for every clock variable $x\in X$,
\item[delay transition] $\langle\ell,\overline{v_i}\rangle\xrightarrow{\delta}\langle\ell,\overline{v_j}\rangle\in T$ for a delay $\delta\in\mathbb{R}_{\geq 0}$, where $\overline{v_j}$ is a clock valuation obtained by incrementing each clock $x\in X$ value by $\delta$: $\overline{v_j}[x]=\overline{v_i}[x]+\delta$ and the location invariant is satisfied $\overline{v_j}\models \mathit{I}(\ell)$,
\item[action transition] $\langle\ell_i,\overline{v_i}\rangle\xrightarrow{a}\langle\ell_j,\overline{v_j}\rangle\in T$ for an edge $(\ell_i,g,a,u,\ell_j)\in E$, clock valuations $\overline{v_i}$ satisfy $g$, and  target valuation $\overline{v_j}$, obtained by applying edge reset $u[\overline{v_i}]$, satisfies $\mathit{I}(\ell_j)$.
\end{description}

A \uppaal TA is extended with discrete variables and C-like functions to ease modeling complex systems. %
Tests (resp. updates) on discrete variables and functions can be used in conjunction with guards (resp. performed together with clock resets). %
\uppaal TA can be composed in parallel into a \emph{network} $\mathit{||_{i\in 1\isep n} A_i}$, $n\in \mathbb{N}_{>1}$. %
\uppaal uses \emph{channels} to allow different $A_i$ to synchronize on actions. 
An exclamation mark (!) denotes the \emph{emitter} and question marks (?) \emph{receivers}. %
In a \emph{handshake} (resp. \emph{broadcast}) channel, an emitter synchronizes with one receiver at a time in a blocking manner (resp. with as many receivers as available; if none, it proceeds alone). %
We refer in this paper to broadcast channels without receivers as \emph{singleton channels}. %
Channels \emph{priorities} may be used to prioritize one concurrent transition over another. %
\emph{Committed locations} are useful to model atomic action sequences. %
A state (in the TLTS) comprising at least one  committed location must be left immediately (no delay transition possible) and the transition taken to leave it must include an edge emanating from a committed location. %

The underlying TLTS of a TA network being typically infinite, \uppaal uses symbolic semantics where each state is represented using a vector of locations (and discrete-variable values), and a symbolic \emph{zone}, defined by a conjunction of constraints of the form $G$. %
\uppaal uses Difference Bound Matrices (DBMs) to maintain the constraint system in a compact canonical form by storing $c_{i,j}$ values in a matrix. %
The matrix values are extended with a special value $\infty$: $c_{i,j}\in\mathbb{Z}\cup\{\infty\}$ meaning that a particular constraint over $x_i-x_j\prec c_{i,j}$ is  absent. %
In practice, removing upper constraints $\{x_i-x_0\leq c_{i,0}\}$ is achieved by setting the corresponding entries in the matrix to the special value $c_{i,0}:=\infty$ and resetting to zero is achieved by assignments $c_{i,0}:=0$ and $c_{0,i}:=0$, and the constraints are propagated by computing a canonical form. %
Details on TA symbolic reachability algorithms are published in~\cite{LLPY97,bengtsson2003timed}. %

\subsubsection{Query Language}\label{sec:queryLanguage}

A state formula is a propositional formula over locations and discrete and clock variables, to be evaluated on a symbolic state. %
For example, $\mathit{A.l\, \, and\, \, A.x \geq 2\, \, and\, \, A.x \leq 3}$ is a state formula that evaluates to true in all symbolic states where $A$ is at location $l$ and the value of clock $x$ in $A$ is comprised between $2$ and $3$. %
In this paper, we use a compact notation when possible, \eg we write the above formula as $\mathit{A.l\, \, and\, \, 2 \leq  A.x \leq 3}$. %
\uppaal supports a subset of CTL to query properties, written as state formulae quantified over using \emph{path formulae}, on TA networks. %
In particular, \uppaal supports $\mathit{sup}$  and $\mathit{inf}$ queries for a specified set of clocks $\{x_i\}$ and a state formula $F$, which explore the entire state space using symbolic constraint solving, record the constraint bounds $c_{i,0}$ (resp. $-c_{0,i}$), and report the maximum (resp. minimum) value of clocks $\{x_i\}$ when $F$ holds. %
For example $\mathit{sup\{A_1.l\, \, and\, \, A_1.x \leq 2\}: A_2.x}$ reports the maximum value of clock $x$ of $\mathit{A_2}$ observed when $\mathit{A_1}$ is in location $l$ and the value of its clock $\mathit{x}$ is less than 2. %
Concrete examples are given in Sect.~\ref{appr}. %

\subsection{RTS Model}\label{rts}
We present a simpler version of FHZ's model for a schedulable RTS, \ie an affinity that guarantees the schedulability of all tasks is known beforehand (obtained \eg using the 
approach in FHZ~\cite[Section 5]{DBLP:journals/jsa/FoughaliHZ23}). 

\subsubsection{Syntax}\label{rts-synt}
An RTS $R$ is a tuple $\mathit{\langle T,C\rangle}$ with $T$ a set of \emph{periodic} tasks and $C$ a set of cores. %

Each $\tau\in T$ is  
a tuple $\mathit{\langle FSM_\tau, P_\tau, \pi_\tau, \affinity_\tau\rangle}$ where $P_\tau \in \mathbb{N}_{>0}$ is $\tau$'s \emph{period} (also its deadline), $\pi_\tau \in \mathbb{N}$ is $\tau$'s \emph{fixed priority}, 
and $\affinity_\tau \in C$ is $\tau$'s affinity, \ie the only core it can execute on (\emph{partitioned  allocation}). %
Each core $c \in C$ is associated with a \emph{partition} $\mathit{\partition_c = \{\tau \in T | \affinity_{\tau} = c\}}$, the affinity's dual. %
$\mathit{FSM_\tau}$ is a finite-state machine $\mathit{\langle S_\tau, act, tr_\tau\rangle}$ representing $\tau$'s behavior where:

\begin{itemize}
\item{}  $\mathit{S_\tau = JS_\tau \cup \{act, end\}}$ is the set of states, with $\mathit{act}$ and $\mathit{end}$ the special \emph{activation} and \emph{termination} states and $\mathit{JS_\tau}$ the set of \emph{segments},
\item{} $\mathit{act}$ is the initial state,
\item{} $tr_\tau \subset S_\tau \times S_\tau$ is the transition relation satisfying (i) $\mathit{act}$ (resp. $\mathit{end}$) has no predecessors (resp. successors) and (ii)  
each \emph{maximal path} from $\mathit{act}$ to $\mathit{end}$ is finite.
\end{itemize}

We obtain accordingly $\mathit{J_\tau}$, the set of \emph{jobs} in $\tau$, mapping each maximal path to the ordered set of segments appearing in it. %

At a lower level, each segment $s$ in the set $\mathit{\mathcal{JS} = \bigcup_{\tau \in T} JS_\tau}$,  is associated with $\mathit{bt_s \in \mathbb{N}_{>0}}$ (resp. $\mathit{wt_s \in \mathbb{N}_{>0}}$), its best-case execution time (BCET) (resp. worst-case execution time (WCET)). %

Two important remarks are worth emphasizing. %
First, the overheads due to data sharing are taken into account, \eg following the approach in FHZ~\cite[Sect. 5]{DBLP:journals/jsa/FoughaliHZ23}. 
That is, for any segment $s$, $\mathit{wt_s}$ includes \emph{blocking} due to data sharing, if any. %
Second, the model above is more expressive than many standard periodic models in the literature. %
For instance, allowing multiple paths within a task enables handling control flows, therefore widening the model's suitability to \eg robotic applications~\cite[Sect. 3]{DBLP:journals/jsa/FoughaliHZ23}.

Fig.~\ref{fig1} (top) illustrates a schedulable RTS with four tasks $\mathit{\tau_1 = \langle FSM_{\tau_1}, 20, 1, c_1\rangle}$ and $\mathit{\tau_2 = \langle FSM_{\tau_2}, 30, 0, c_1\rangle}$ (allocated to core $c_1$), and $\mathit{\tau_3 = \langle FSM_{\tau_3}, 20, 1, c_2\rangle}$ and $\mathit{\tau_4 = \langle FSM_{\tau_4}, 40, 0, c_2\rangle}$ (allocated to $c_2$); the timing constraints of a segment $s$ are indicated on the corresponding state 
using a couple $\mathit{(bt_s, wt_s)}$. %
For instance, $\tau_2$ contains three segments $\mathit{JS_{\tau_2} = \{s2, s3, s4\}}$, with $\mathit{bt_{s2} = 1}$, $\mathit{bt_{s3} = 3}$, $\mathit{bt_{s4} = 2}$, $\mathit{wt_{s2} = 3}$, $\mathit{wt_{s3} = 6}$, $\mathit{wt_{s4} = 5}$, and three jobs $\mathit{J_{\tau_2} = \{ \{s_2, s_3\}, \{s_4, s_3\}, \{s_4\} \}}$. %
This RTS example will be used throughout the paper. %  

The semantics of an RTS, thoroughly discussed in FHZ, 
relies on P-FP scheduling with limited preemption and is exemplified using \uppaal models next.

\subsubsection{\uppaal Model}\label{rts-upp} 
\begin{figure}[tb!]
\centering
\includegraphics[width=0.9\columnwidth]{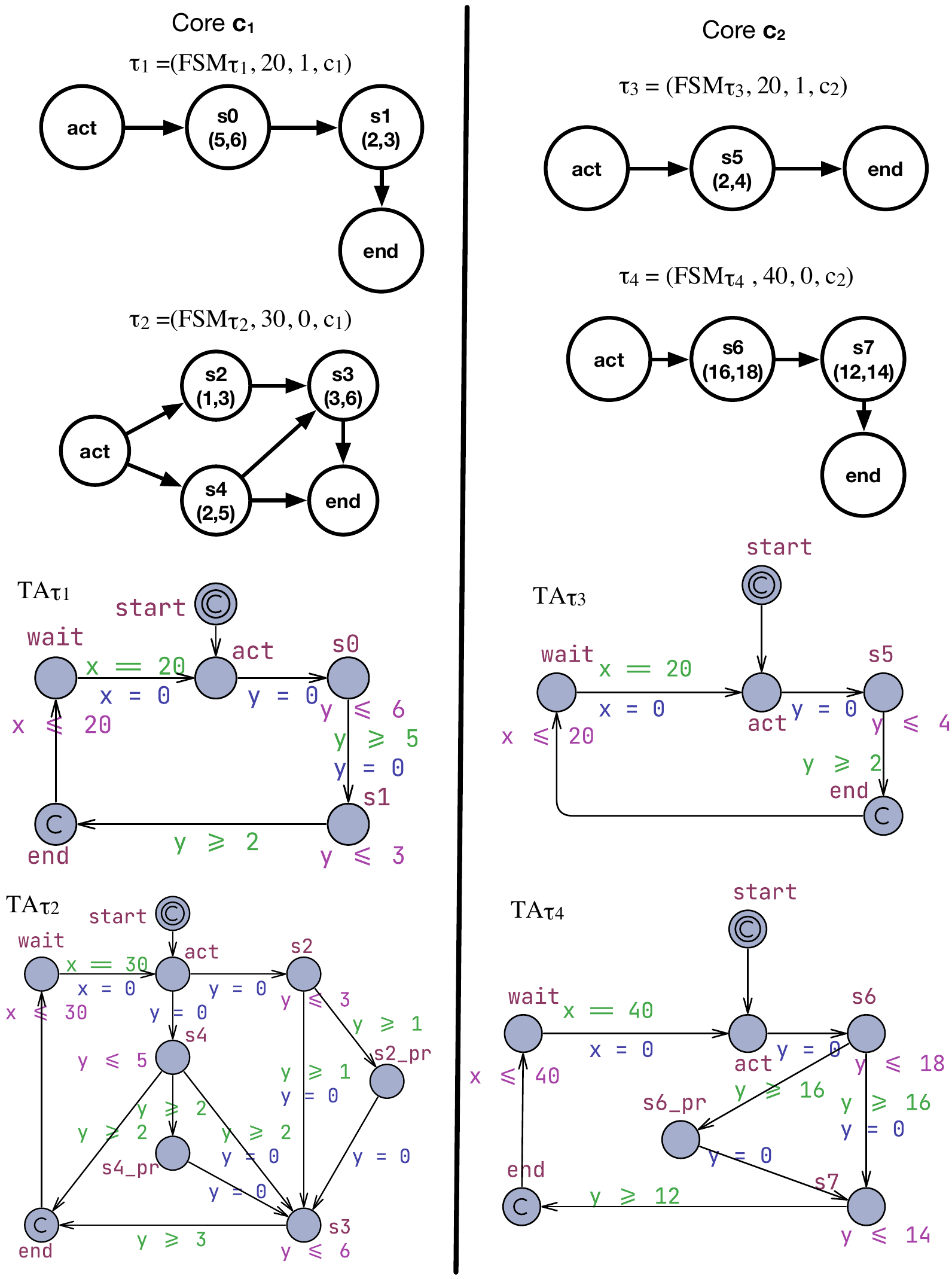}
\caption{An example of four tasks and their \uppaal models.}

\label{fig1}
\end{figure}

\paragraph{Tasks.} 
FHZ provide definitions to generate a \uppaal TA $\mathit{TA_\tau}$ from any task $\tau$ with the above syntax~\cite[Sect. 6]{DBLP:journals/jsa/FoughaliHZ23}. %
We illustrate what such definitions give for the RTS example in Fig~\ref{fig1} (top). %
Fig~\ref{fig1} (bottom) depicts accordingly $\mathit{TA_\tau}$ of tasks $\tau \in \{\tau_1, \tau_2, \tau_3, \tau_4\}$. %
We explain using $\mathit{TA_{\tau_2}}$. %
We use the notation $s\to s'$ to denote an edge from $s$ to $s'$, and $\to s$ (resp. $s\to$) to denote all incoming (resp. outgoing) edges of $s$. %

$\tau_2$ is \emph{activated} at each multiple of its period (multiples of $30$). %
This is ensured, in $\mathit{TA_{\tau_2}}$, through the invariant at $\mathit{wait}$, and the guard and reset of clock $x$ on $\mathit{wait\to act}$. %
The activation at $0$ is handled through the initial, committed location $\mathit{start}$. %
Edges $\mathit{act\to}$ correspond to \emph{releasing} $\tau_2$, \ie allowing it to execute. %
Since time until such release is unknown, $\mathit{act}$ is invariant free. %
Executing a job in $\tau_2$ corresponds to traversing a path in $\mathit{TA_{\tau_2}}$ from $\mathit{act}$ to $\mathit{end}$. %
Executing a segment $s$ between $\mathit{bt_s}$ and $\mathit{wt_s}$ is ensured by the invariant at location $s$, the guards on edges $s\to$, and the resets of clock $y$ on edges $\to s$.  %
Preemption may happen at the end of a segment, by taking an edge $\mathit{s\to s_{pr}}$. %
Since the time until $\tau_2$ is next released (after being preempted) is unknown, locations $\mathit{s_{pr}}$ are invariant free. %
Notice the absence of preemption location $\mathit{s3_{pr}}$ %
(since $\tau_2$ can only terminate after executing $\mathit{s3}$, its preemption at the end of $\mathit{s3}$ coincides with its termination). %
On the other tasks TA, notice how \eg % 
$\mathit{TA_{\tau_1}}$ does not contain preemption locations (since $\tau_1$ has the highest priority among tasks allocated to $c_1$). %
Also, since tasks are schedulable, it is impossible to reach location $\mathit{end}$ with the value of clock $x$ larger than $P_\tau$ in some $TA_\tau$. %
That is, locations used to handle deadline violation in FHZ are eliminated %
 after verifying that the RTS is schedulable. %

\begin{figure*}[tb!]%
\centering
\includegraphics[width=0.99\textwidth]{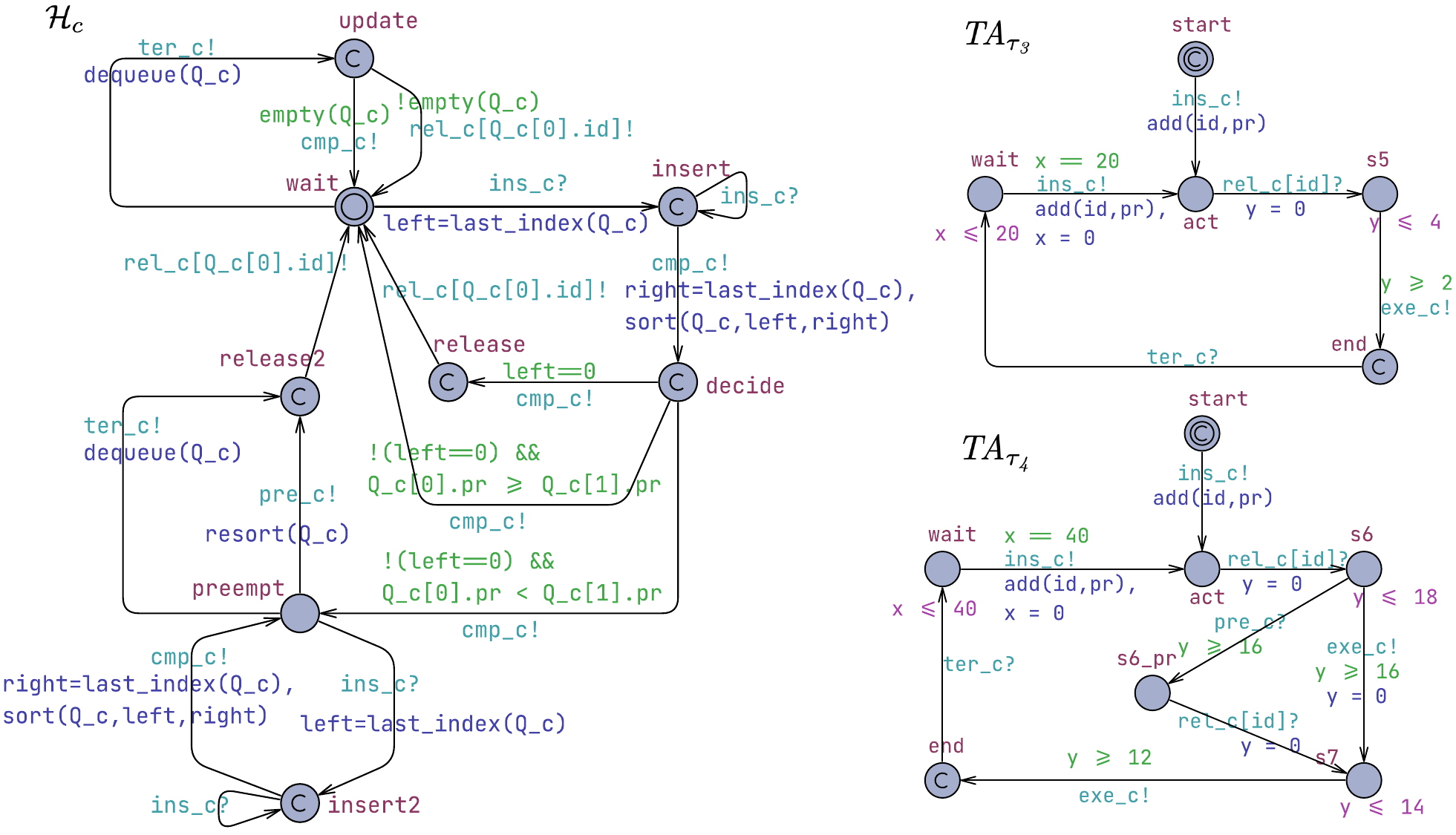}
\caption{Network $N_{c_2}$ corresponding to core $c_2$ from Fig.~\ref{fig1}. To simplify notations, we rename $c_2$ as $c$.} 
\label{fig2}%
\end{figure*}

\paragraph{RTS.} 
$\mathit{TA_\tau}$, for each $\tau$ allocated to some core $c$, are composed with $H_c$, the (P-FP with limited preemption) scheduler of $c$, to obtain $\mathit{N_c = H_c ||_{\tau\in{\partition_c}} TA_\tau}$, the network of core $c$. %
The RTS network is therefore $\mathit{N = ||_{c\in C} N_c}$.  %
In $N_c$, the behavior of tasks is influenced by the scheduler $H_c$ as their respective TA synchronize through channels and variables (see below), whereas in $N$, each $N_c$ is independent (since blocking, if any, is integrated in segments WCETs). %
 
We explain the behavior of $\mathit{N_{c_2} = H_{c_2} || TA_{\tau_3} || TA_{\tau_4}}$ (Fig.~\ref{fig2}), the TA network corresponding to core $c_2$ and tasks $\tau_3$ and $\tau_4$ in Fig.~\ref{fig1}. %
To simplify notations, we rename $c_2$ as $c$. %
We analyze the behaviors of $\mathit{N_{c}}$ between times $0$ and $40$. %
At time $0$, $H_c$ synchronizes, on handshake channel $\mathit{ins_c}$, with $\mathit{TA_{\tau_3}}$ and $\mathit{TA_{\tau_4}}$ atomically one after another. %
This is ensured through committed location $\mathit{insert}$ and the priority $\mathit{ins_c > cmp_c}$ ($\mathit{cmp_c}$ is a singleton channel). %
At each synchronization on $\mathit{ins_c}$, a task inserts ``itself'', \ie its unique identifier $\mathit{id}$ and its priority $\mathit{pr}$ in $Q_c$, a queue of records of two integers, using the function $\mathit{add(id,pr)}$. %
From location $\mathit{insert}$, $H_c$ sorts all tasks in $Q_c$ according to their priorities (to their order of arrival if priorities are equal) and reaches committed location $\mathit{decide}$. %
This sorting does not involve the task at $Q_c[0]$ (the head of $Q_c$) if it is already executing. %
Here, $Q_c$ was empty before the insertion started, so $\mathit{TA_{\tau_3}}$ is now at $Q_c[0]$. %
$H_c$ immediately releases $\mathit{TA_{\tau_3}}$, using its $\mathit{id}$, through the handshake channel $\mathit{rel_c[id]}$ (edge $\mathit{decide\to release}$ then $\mathit{release\to wait}$). % 
$\mathit{TA_{\tau_3}}$ then executes $\mathit{s5}$ and reaches committed location $\mathit{end}$ at a time between $2$ and $4$. %
Then, it atomically synchronizes with $H_c$ (handshake channel $\mathit{ter_c}$) as the latter removes it from $Q_c$ (function $\mathit{dequeue}$, edge $\mathit{wait\to update}$), then immediately releases $\mathit{TA_{\tau_4}}$ which starts executing $\mathit{s6}$ accordingly. %
Such execution ends between times $2+16 = 18$ and $4+18 = 22$.

If $\mathit{s6}$ is left between $18$ and $20$, $\mathit{TA_{\tau_4}}$ moves to $\mathit{s7}$ immediately, %
the execution of which ends between times $18 + 12 = 30$ and $20 + 14 = 34$. %
Meanwhile, at time $20$, $\mathit{TA_{\tau_3}}$ is activated again, $H_c$ moves to location $\mathit{insert}$ and $\mathit{TA_{\tau_3}}$ to location $\mathit{act}$. %
Now, $H_c$ sorts only tasks behind $\mathit{TA_{\tau_4}}$, already executing, in $Q_c$. %
Then, from location $\mathit{decide}$, since $\pi_{\tau_4} < \pi_{\tau_3}$, $\mathit{TA_{\tau_4}}$ must be preempted; $H_c$ moves immediately to location $\mathit{preempt}$. %
Since the only successor of $\mathit{s7}$ is $\mathit{end}$, the preemption of $\mathit{TA_{\tau_4}}$ coincides with its termination. %
Therefore, between times $30$ and $34$, $\mathit{TA_{\tau_4}}$ synchronizes on $\mathit{ter_c}$ with $H_c$ as the latter removes it from $Q_c$ %
and moves to committed location $\mathit{release2}$. %
Immediately, $H_c$ releases $\mathit{TA_{\tau_3}}$, now at $Q_c[0]$, and transits back to $\mathit{wait}$. %
$\mathit{TA_{\tau_3}}$ then executes $\mathit{s5}$ between times $30 + 2 = 32$ and $34 + 4 = 38$ and terminates. 

If, instead, $\mathit{s6}$ is left between $20$ and $22$, $\mathit{TA_{\tau_4}}$ is forced to preempt while at location $\mathit{s6}$ (edge  $\mathit{s6\to s6_{pr}}$) %
as $H_c$ transits to $\mathit{release2}$, synchronizing on channel $\mathit{pre_c}$ (ensured by the priority $\mathit{pre_c > exe_c}$, $\mathit{exe_c}$ is a singleton channel). %
Together with this transition, $H_c$ re-sorts $Q_c$ by putting $\mathit{TA_{\tau_4}}$ behind $\mathit{TA_{\tau_3}}$ (function $\mathit{resort}$) and releases $\mathit{TA_{\tau_3}}$,  %
which then executes $\mathit{s5}$ between times $20 + 2 = 22$ and $22 + 4 = 26$ and terminates. % 
$H_c$ then removes $\mathit{TA_{\tau_3}}$ from $Q_c$ (transitioning to $\mathit{update}$) then immediately releases $\mathit{TA_{\tau_4}}$ (transitioning back to $\mathit{wait}$) which will accordingly resume its execution at $\mathit{s7}$ (moving from $\mathit{s6_{pr}}$ to $\mathit{s7}$). %
$\mathit{TA_{\tau_4}}$ terminates then between $22 + 12 = 34$ and $26 + 14 = 40$. %

We conclude with three remarks. %
First, FHZ explores all possible orderings of simultaneous events. %
For instance, if $\mathit{TA_{\tau_4}}$ finishes executing $\mathit{s6}$ at the same time $\mathit{TA_{\tau_3}}$ is activated (\ie $20$), the subsequent behavior depends on which of these two events happened \emph{first} at this particular instant. This explains why this border case belongs to both scenarios above. %
Second, the description above is example-oriented and high level, whereas $H_{c}$ follows in fact a universal model for P-FP scheduling with limited preemption (we refer the interested reader to~\cite[Sect. 6]{DBLP:journals/jsa/FoughaliHZ23} for   low-level details). %
Finally, the set of behaviors in our example between times $40$ and $80$, $80$ and $120$, etc. is identical to the one depicted between times $0$ and $40$. %
This is because, in a schedulable network $N_c$, the behaviors repeat at each \emph{hyperperiod} $\mathit{hp_c}$, given by the least common multiple of periods of tasks allocated to $c$, \ie $\mathit{hp_c = lcm(\{P_\tau | \tau\in\partition_c\})}$. %

\section{Approach}\label{appr}
In this section, we present the core of our contribution. %
Note that the presentation throughout this section, including the used examples and the proposed algorithms, favors simplicity over technical details. %
The latter are discussed in Sect.~\ref{impl}. %

\subsection{Problem Statement \& Motivation}\label{not} Let $R = \langle T, C\rangle$ be a schedulable RTS, $\mathit{N = ||_{c\in C} N_c}$ its \uppaal model, and $E$ a non-empty set of events of interest. % 
Let $\mathit{T_{E} \subseteq T}$ be the smallest subset of tasks producing events in $E$. %
A task $\tau$ in $T_E$ produces events through segments in the set $\mathit{JS^{E}_\tau}$. %
For simplicity, we assume that such segments belong to different jobs in $J_\tau$, \ie $\mathit{\forall s,s'\in JS^{E}_\tau, J \in J_\tau: s\in J \wedge s'\in J \Rightarrow s = s'}$. %
The set of all event-producing segments in $R$ is therefore $\mathit{\mathcal{JS}_E = \cup_{\tau\in T_E} JS^{E}_\tau}$. 
Let $\mathit{C_{E} \subseteq C}$ be the smallest subset of cores covering the affinities of tasks in $\mathit{T_{E}}$ ($\mathit{C_{E} = \{ c \in C | T_{E} \cap \partition_c\neq \varnothing\}}$).  %
Events occur on at least two cores, \ie $\mathit{\card{C_{E}} \geq 2}$. %
We require that tasks in $T_E$ have \emph{distinct affinities}, \ie  $\card{T_E} = \card{C_E}$. % 
The reason behind this restriction is detailed in Sect.~\ref{distaff}. %
If $\mathit{s \in JS^{E}_\tau}$ produces event $e$, we say that $\tau$ produces $e$, and that $e$ occurs on $c = \affinity_\tau$. %
There is no restriction on the number of cores an event can occur on. %

A segment $\mathit{s}$ in $\mathit{\mathcal{JS}_E}$ may produce a sequence of an arbitrary number of events. %
This is handy in practice, \eg when a segment reads and writes some data (Sect.~\ref{eval}). %
Let $\mathbb{I}_s$ be the set of non-empty closed intervals of reals with natural bounds within the relative execution time of segment $s\in \mathit{\mathcal{JS}_E}$. %
That is, $[a,b] \in \mathbb{I}_s$, $\mathit{a,b\in \mathbb{N}, a\leq b \leq wt_s}$, equals the set of reals $\{x \in \mathbb{R} | \mathit{a\leq x\leq b}\}$. %
We associate each segment $s$ in $\mathit{\mathcal{JS}_E}$ with a set of couples $\mathit{s_E \subset E \times \mathbb{I}_s}$. %
Each $\mathit{i^{th}}$ element of $s_E$ represents an event $e$ that $s$ produces as well as the interval within which $s$ may produce $e$, relative to the execution of $s$. %
For example, if $s$ produces one event $e$ at the end of its execution then $\mathit{s_E = \{(e, [bt_s,wt_s])\}}$. %
The set $s_E$ is \emph{ordered}. %
An event belonging to a couple at position $i$ of $s_E$ occurs \emph{always} before an event appearing in a couple at position $j > i$. %
For intervals, we impose an order on their bounds: any two elements $\mathit{[a,b]}$ and $\mathit{[a',b']}$ appearing in couples at positions $i$ and $j$, $j>i$, respectively, obey the inequalities $a\leq a'$ and $b\leq b'$. %
That is, intervals may overlap, but they must respect the strict ordering of events occurrences. %
For instance, if $\mathit{s_E = \{(e,[a,b]),(e',[a',b'])\}}$, and $\mathit{[a,b] \cap [a',b'] = [c,d] \neq \varnothing}$, then $e$ still occurs before $e'$ in the interval $\mathit{[c,d]}$ (\ie the possible occurrence of $e'$ in $\mathit{[c,d]}$ is conditioned by a prior occurrence of $e$). In brief, we remain the most generic possible as we impose no constraints on \emph{when} a segment produces an event within its execution, since elements of $\mathbb{I}_s$ can be any interval included in $[0,wt_s]$. %

The goal is to compute \emph{exact} inter-core bounds following some requirement. %
For example, given $\mathit{E = \{e_1, e_2, e_3\}}$, what is the exact maximum amount of time separating each production of $e_1$ and the next production of $e_2$, with $e_3$ happening in between? %
One motivation behind exactness is the fact that existing RTS (multicore) models, on which analyses are carried out, typically include over-approximations due to blocking. %
FHZ, for instance, integrate blocking in segments WCETs using an analytical approach. %
Though the latter is particularly tight, the resulting WCETs may be slightly over-approximated~\cite[Sect. 5]{DBLP:journals/jsa/FoughaliHZ23}. %   
Consequently, it is important to be exact \wrt to a model  of $R$ (in our case $N$) that inevitably over-approximates the effects of blocking in one way or another, therefore avoiding further pessimism. %
The \emph{direct method} to achieve this is to compose $\mathit{N_{E} = ||_{c\in C_{E}} N_c}$ with an \emph{observer} $\mathit{Obs}$, as suggested in FHZ. % 
However, given the complexity of each $N_c$, building the state-space of $\mathit{N_{E} || Obs}$ does not scale for large systems. %
For example, if we try to compute an inter-core bound on the 2017 WATERS industrial challenge (for which FHZ successfully computed WCRTs on one core), the direct method leads to memory exhaustion (Sect.~\ref{eval}). %
We need therefore to use \emph{abstractions} to reduce the state-space size.

\subsection{Core Idea}\label{ci} %
Our main idea is the following. %
As long as building the state space of each $N_c$ apart is scalable, we use this to our advantage as we  compute a (much smaller) abstraction $\mathcal{A}(N_c)$ that preserves the \emph{absolute intervals} within which events occur on $c$, \ie produced by the only task, say $\tau$, in $\mathit{T_E\cap \partition_c}$. %
This abstraction is based on the original $N_c$ behavior, and therefore takes into account all the delays event-producing segments in $\tau$ may incur due to other tasks allocated to $c$, \ie in $\mathit{\partition_c\backslash\{\tau\}}$ (and blocking due to tasks allocated to other cores, already integrated in $N_c$). %
Then, instead of computing the bounds on $\mathit{N_{E} || Obs}$, we do so on $\mathit{\mathcal{A}(N_{E}) || Obs}$, where $\mathit{\mathcal{A}(N_{E}) = ||_{c\in C_{E}} \mathcal{A}(N_c)}$, therefore drastically reducing the state space. %
We first present a coarse abstraction and explain its weakness (Sect.~\ref{ca}). %
Then, we devise an exact abstraction, relying on a new query that we implement in \uppaal  (Sect.~\ref{ea}). %

\subsection{Examples} For illustration, we make use of simple\footnote{We demonstrate our approach's scalability on the real WATERS 2017 industrial challenge in Sect.~\ref{eval}.} examples, all relying on the schedulable RTS in Sect.~\ref{rts} (Figures.~\ref{fig1}, ~\ref{fig2}). %
The hyperperiods are $\mathit{hp_{c_1} = 60}$ and $\mathit{hp_{c_2} = 40}$. %

\subsubsection{Example 1}\label{ex1}
$\mathit{E = \{e_1, e_2\}}$ with $e_1$ produced by segment $\mathit{s5}$ and $e_2$ by $\mathit{s1}$. % 
Both events are produced at the end of execution of such segments, that is at any instant comprised between their best- and worst-case execution times. %
Therefore, $\mathit{C_{E} = \{c_1, c_2\}}$, $\mathit{T_{E} = \{\tau_1, \tau_3\}}$, $\mathit{JS^{E}_{\tau_1} = \{s1\}}$, $\mathit{JS^{E}_{\tau_3} = \{s5\}}$, $\mathit{s5_E = \{(e_1,[2,4])\}}$ and $\mathit{s1_E = \{(e_2,[2,3])\}}$. %

\subsubsection{Example 2}\label{ex2} Example 1 with a third event $e_3$ produced by $\mathit{s5}$ in the relative interval $[0,1]$. %
$\mathit{E = \{e_1, e_2, e_3\}}$, $\mathit{C_{E} = \{c_1, c_2\}}$, $\mathit{T_{E} = \{\tau_1, \tau_3\}}$, $\mathit{JS^{E}_{\tau_1} = \{s1\}}$, $\mathit{JS^{E}_{\tau_3} = \{s5\}}$, $\mathit{s5_E = \{(e_3,[0,1]),(e_1,[2,4])\}}$, and $\mathit{s1_E = \{(e_2,[2,3])\}}$.

\subsubsection{Example 3}\label{ex3} Example 2 with $\tau_1$ no longer producing events. Instead, $\tau_2$ produces event $e_4$ (resp. $e_2$) through $\mathit{s2}$ in the relative interval $[0,3]$ (resp. $\mathit{s4}$ in the relative interval $[2,4]$). %
$\mathit{E = \{e_1, e_2, e_3, e_4\}}$, $\mathit{C_{E} = \{c_1, c_2\}}$, $\mathit{T_{E} = \{\tau_3, \tau_2\}}$, $\mathit{JS^{E}_{\tau_2} = \{s2,s4\}}$, $\mathit{JS^{E}_{\tau_3} = \{s5\}}$, $\mathit{s5_E = \{(e_3,[0,1]),(e_1,[2,4])\}}$, $\mathit{s2_E = \{(e_4,[0,3])\}}$ and $\mathit{s4_E = \{(e_2,[2,4])\}}$.

\subsection{Coarse Abstractions}\label{ca} 

Consider Example 1 (Sect.~\ref{ex1}). %
We want to compute $\mathit{B^\mathit{max}_{e_1, e_2}}$, the \emph{maximal} inter-core bound between each production of $e_1$ and 
 the next production of $e_2$. %  

\begin{figure}[tb!]%
\centering
\includegraphics[width=0.8\columnwidth]{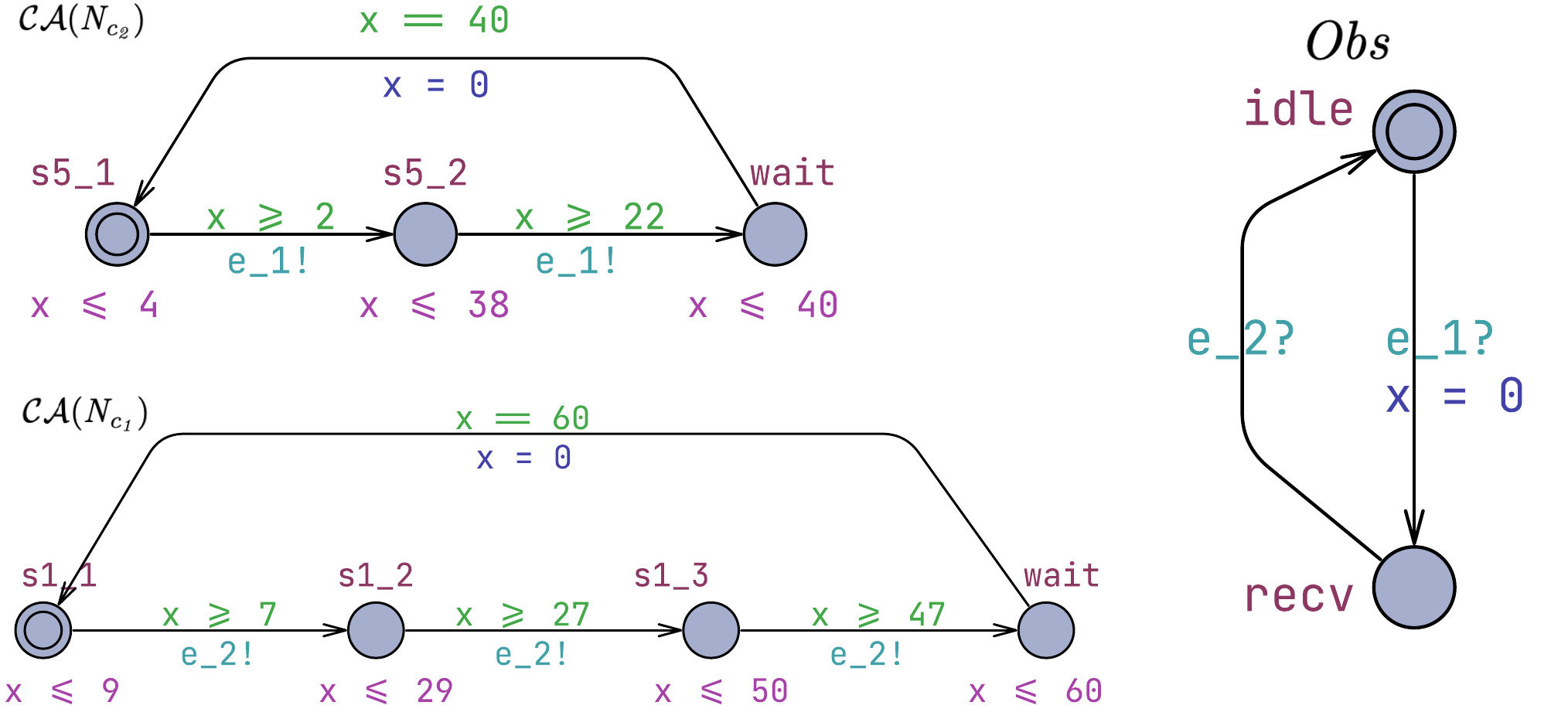}
\centering
\caption{Coarse abstractions for cores $c_1$ (bottom left) and $c_2$ (top left) and observer $\mathit{Obs}$ (right), for Example 1 (Sect.~\ref{ex1}).} \label{fig3}%
\end{figure}

For $N_{c_1}$, we build a coarse abstraction $\mathit{\mathcal{CA}(N_{c_1})}$ as follows:
\begin{itemize}
\item{} Add a clock $x$ to $H_{c_1}$, 
\item{} Run, on $N_{c_1}$ separately, \uppaal with the queries: \\
$\mathit{min_{s1,e_2,1} = inf\{\phi_{s1, e_2}\, \, and\, \, H_{c_1}.x \leq 20\}:  H_{c_1}.x}$\\
$\mathit{max_{s1,e_2,1} = sup\{\phi_{s1, e_2}\, \, and\, \, H_{c_1}.x \leq 20\}:  H_{c_1}.x}$\\
$\mathit{min_{s1,e_2,2} = inf\{\phi_{s1, e_2}\, \, and\, \, 20 \leq H_{c_1}.x \leq 40\}:  H_{c_1}.x}$\\
$\mathit{max_{s1,e_2,2} = sup\{\phi_{s1, e_2}\, \, and\, \, 20 \leq H_{c_1}.x \leq 40\}:  H_{c_1}.x}$\\
$\mathit{min_{s1,e_2,3} = inf\{\phi_{s1, e_2}\, \, and\, \, 40 \leq H_{c_1}.x \leq 60\}:  H_{c_1}.x}$\\
$\mathit{max_{s1,e_2,3} = sup\{\phi_{s1, e_2}\, \, and\, \, 40 \leq H_{c_1}.x \leq 60\}:  H_{c_1}.x}$\\
with $\mathit{\phi_{s1, e_2} = TA_{\tau_1}.s1\, \, and\, \, 2 \leq TA_{\tau_1}.y \leq 3}$,
\item{} Obtain $\mathit{min_{s1,e_2,1} = 7}$, $\mathit{max_{s1,e_2,1} = 9}$, $\mathit{min_{s1,e_2,2} = 27}$, $\mathit{max_{s1,e_2,2} = 29}$, \\ $\mathit{min_{s1,e_2,3} = 47}$  and $\mathit{max_{s1,e_2,3} = 50}$,
\item{} Build $\mathit{\mathcal{CA}(N_{c_1})}$ accordingly (Fig.~\ref{fig3}, bottom left).
\end{itemize}

And for $N_{c_2}$:

\begin{itemize}
\item{} Add a clock $x$ to $H_{c_2}$, 
\item{} Run, on $N_{c_2}$ separately, \uppaal with the queries: \\
$\mathit{min_{s5,e_1,1} = inf\{\phi_{s5, e_1} \, \, and\, \, H_{c_2}.x \leq 20\}:  H_{c_2}.x}$\\
$\mathit{max_{s5,e_1,1} = sup\{\phi_{s5, e_1} \, \, and\, \, H_{c_2}.x \leq 20\}:  H_{c_2}.x}$\\
$\mathit{min_{s5,e_1,2} = inf\{\phi_{s5, e_1} \, \, and\, \, 20\leq H_{c_2}.x \leq 40\}:  H_{c_2}.x}$\\
$\mathit{max_{s5,e_1,2} = sup\{\phi_{s5, e_1} \, \, and\, \, 20 \leq H_{c_2}.x \leq 40\}:  H_{c_2}.x}$\\
with $\mathit{\phi_{s5, e_1} = TA_{\tau_3}.s5\, \, and \, \,  2 \leq TA_{\tau_3}.y \leq 4}$,
\item{} Obtain $\mathit{min_{s5,e_1,1} = 2}$, $\mathit{max_{s5,e_1,1} = 4}$, $\mathit{min_{s5,e_1,2} = 22}$, and $\mathit{max_{s5,e_1,2} = 38}$,
\item{} Build $\mathit{\mathcal{CA}(N_{c_2})}$ accordingly (Fig.~\ref{fig3}, top left).
\end{itemize}

Let us explain how this works, using the example of core $c_2$. %
Adding clock $x$ to the scheduler allows to have an \emph{absolute reference} of time (this clock is never reset, and all clocks start synchronously at $0$ in a TA network, Sect.~\ref{upp}). %
Relying on this reference, we compute, in each period of $\tau_3$, the task producing $e_1$ through segment $\mathit{s5}$, and up to the hyperperiod $\mathit{hp_{c_2}}$, the exact \emph{earliest} and \emph{latest} production times of $e_1$, using the original (non-abstracted) network
 $N_{c_2}$. % 
For example, $\mathit{max_{s5,e_1,2}}$ stores the latest time at which $e_1$ is produced by $\mathit{s5}$ within the second period of $\tau_3$. %
This corresponds to the largest value of the scheduler's clock $x$ when $\tau_3$ is at location $\mathit{s5}$ and the value of $\tau_3$'s clock $y$ is comprised between $\mathit{bt_{s5}}$ and $\mathit{wt_{s5}}$ (state formula $\phi_{s5, e_1}$) \emph{and} the absolute time is greater than the first multiple of $\mathit{P_{\tau_3}}$ and less than the second multiple of $\mathit{P_{\tau_3}}$ ($\mathit{20 \leq H_{c_2}.x \leq 40}$). %
Once all such times are acquired, we simply abstract $\mathit{N_{c_2}}$ into $\mathit{\mathcal{CA}(N_{c_2})}$ where, in each $\mathit{k^{th}}$ period ($\mathit{k \in 1\isep \frac{hp_{c_2}}{P_{\tau_3}}}$) of $\tau_3$, $e_1$ is produced between $\mathit{min_{s5,e_1,k}}$ and $\mathit{max_{s5,e_1,k}}$ (recall that $\tau_3$ is the only task in $T_E$ with affinity $c_2$). %
When the multiple of the period coincides with the hyperperiod, location $\mathit{wait}$ ensures waiting for the hyperperiod to be reached, after which the behavior repeats by transiting back to the initial location of $\mathit{\mathcal{CA}(N_{c_2})}$ and resetting its clock $x$. % 
Before going any further, we recall that the earliest and latest production times are ensured to be \emph{exact} \wrt the model of the RTS, since we compute them on the \emph{original core network} $N_c$ (Sect.~\ref{ci}). %
In the case of core $c_2$, for instance, we obtain $\mathit{min_{s5,e_1,2} = 22}$ and $\mathit{max_{s5,e_1,2} = 38}$, which coincide with the analysis we carried out on $N_{c_2}$ within the second period of $\tau_3$ (Sect.~\ref{rts-upp}). %

We compose then both core abstractions to obtain a new (smaller) network $\mathit{\mathcal{CA}(N_{E}) = \mathcal{CA}(N_{c_1}) || \mathcal{CA}(N_{c_2})}$, which we compose in parallel with the observer $\mathit{Obs}$ in Fig.~\ref{fig3} (right). %
Here, $\mathit{Obs}$ receives $e_1$ and $e_2$ on broadcast channels having the same name. %
From its initial location $\mathit{idle}$, when an $e_1$ occurs, it  
transits to location $\mathit{recv}$ and resets its clock $x$. %
At location $\mathit{recv}$, it awaits the subsequent occurrence of $e_2$. %
Therefore, to compute the  
maximum % 
inter-core bound % 
$\mathit{B^\mathit{max}_{e_1, e_2}}$ %
separating each $e_1$ and the next $e_2$, it suffices to query % 
$\mathit{sup\{Obs.recv\}:Obs.x}$ % 
on $\mathit{\mathcal{CA}(N_{E}) || Obs}$. %
Note that $\mathit{Obs}$ is the simplest possible to obtain $\mathit{B^\mathit{max}_{e_1, e_2}}$, \eg it is not suitable for computing minimal bounds. %
In Sect.~\ref{eval}, we show how we leverage the power of observers to compute various types of bounds. % 

\begin{figure}[tb!]%
\centering
\includegraphics[width=0.8\columnwidth]{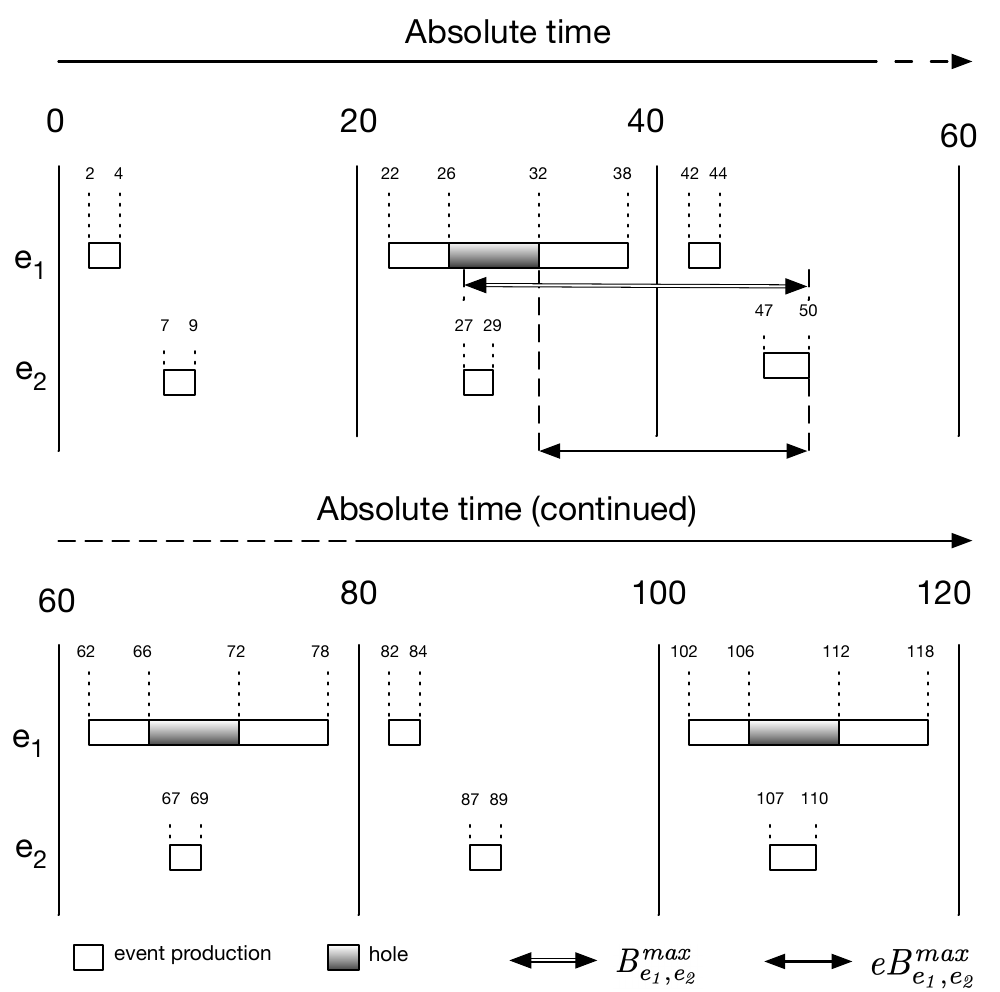}
\centering
\caption{The hole phenomenon (Example 1, Sect.~\ref{ex1}.)} \label{fig4}%
\end{figure}

\paragraph{The Hole Phenomenon.} % 
Unfortunately, inter-core bounds computed using coarse abstractions may be not exact. %
For instance, the procedure above gives % 
$\mathit{B^\mathit{max}_{e_1, e_2} = 23}$. %
This value is in fact over-approximated. %
We illustrate this in Fig.~\ref{fig4}, where we depict all intervals within which $e_1$ and $e_2$ may be produced on a scale of absolute time, up to the least common multiple of both hyperperiods $\mathit{hp_{c_1}}$ and $\mathit{hp_{c_2}}$, that is $\mathit{hp = 120}$. %
The behavior repeats afterwards, \ie in Fig.~\ref{fig4}, when reaching time $120$, the behavior continues by going back to $0$ and shifting the time scale to the right by $\mathit{hp}$, then again when reaching $240$ and so on. %
The white rectangles represent the exact absolute intervals within which an event may occur. %
For $e_1$, they come from the analysis we carried out in Sect.~\ref{rts-upp} (and for $e_2$ through a similar analysis). %
We can easily see that %
 $\mathit{B^\mathit{max}_{e_1, e_2} = 23}$ (double-stroke double-headed arrow), is larger than $\mathit{eB^\mathit{max}_{e_1, e_2} =18}$, the exact one (single-stroke double-headed arrow). %
In the worst case (in the sense of maximizing the amount of time between any $e_1$ and a subsequent  $e_2$), $e_1$ occurs at absolute time $32 + n \cdot hp$ and $e_2$ occurs at $50 + n \cdot hp$, where $n$ is a positive integer. %
For $n = 0$, %
the worst case corresponds to $e_1$ occurring at the very beginning of the second absolute production interval within the second period of $\tau_3$, \ie at exactly $32$. %
Now, the next $e_2$ production will not happen before the third period of $\tau_1$, because $e_2$ has already been produced in its second period between $27$ and $29$. %
Therefore, the subsequent $e_2$ will be produced at $50$ in the worst-case scenario. %
The exact maximal bound is therefore equal to $50 - 32 = 18$. %

The reason behind this weakness of coarse abstractions is the possible existence of \emph{holes}, illustrated using gray rectangles in Fig.~\ref{fig4}. %
In a nutshell, coarse abstractions capture the exact earliest and latest production times of an event within a given period, but do not take into account non-empty intervals in between (holes) within which the event may never occur, \ie both white and gray rectangles are treated as white ones. %
Therefore, in the coarse abstraction shown in Fig.~\ref{fig3}, $e_1$ may occur at any point between $22$ and $38$, including within the hole $\mathit{]26,32[}$. %
Consequently, the worst case scenario corresponds to both events happening at $27$ in the order $e_2$ then $e_1$, which explains the obtained maximal bound $50 - 27 = 23$.   

\subsection{Exact Abstractions}\label{ea} To overcome the hole phenomenon, we devise a new, exact abstraction. %
For this, a new \uppaal query is needed to extract the ``white rectangles'' in the sense of Fig.~\ref{fig4}.  

\subsubsection{A new \uppaal query}\label{new-query}

In a general sense, the new query must be able to compute % 
all the contiguous intervals of values of a clock $x$ within which some state formulae $F$ holds. %
We implemented this query, called $\mathit{bounds}$, in \uppaal 5.1.0. %
In brief, $\mathit{bounds\{F\}: A.x}$ generalizes $\mathit{sup}$ and $\mathit{inf}$ queries (Section~\ref{sec:queryLanguage}) in a sense that it reports both the minimum and maximum value of $x$, and in addition reports a precise union of possible values intervals of $x$ when $F$ holds. %
Such collection of value intervals does not pose a significant overhead besides exploring the full symbolic state space, which is already the case for $\mathit{sup}$ and $\mathit{inf}$ (Sect.~\ref{eval}). %

\subsubsection{Using the bounds Query}\label{uq} 
\begin{figure}[tb!]%
\centering
\includegraphics[width=0.65\columnwidth]{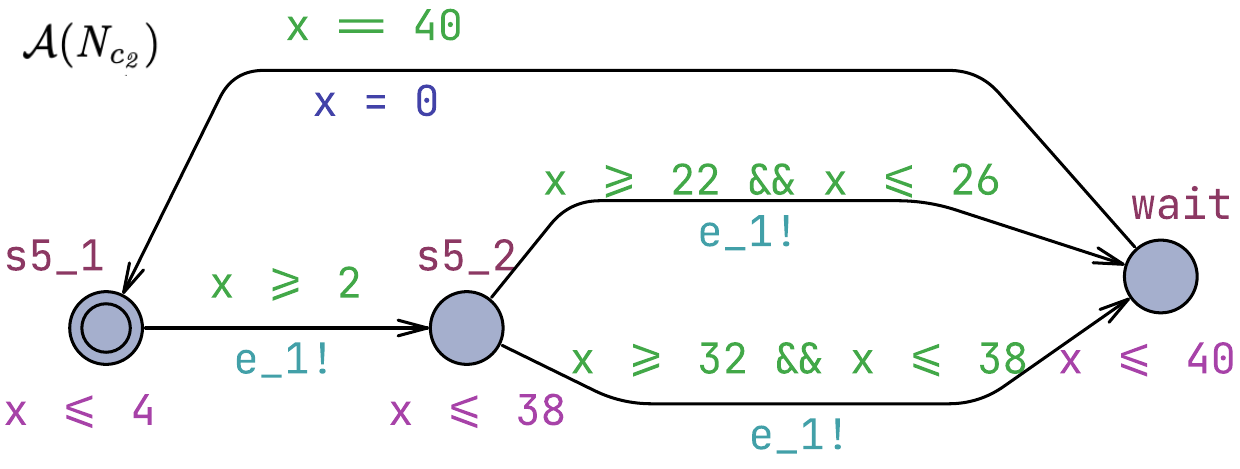}
\centering
\caption{Exact abstraction of core $c_2$ % 
for Example 1 (Sect.~\ref{ex1}). %
The exact abstraction of core $c_1$ is the same as its coarse abstraction in Fig.~\ref{fig3}.}    \label{fig5}%
\end{figure}

Back to Example 1 (Sect.~\ref{ex1}). %
To compute exact abstractions $\mathit{\mathcal{A}(N_{c_2})}$ and $\mathit{\mathcal{A}(N_{c_1})}$, we use a similar procedure as for coarse abstractions, with two main differences. %
First, instead of querying with $\mathit{inf}$ and $\mathit{sup}$, we query with $\mathit{bounds}$ to compute the exact absolute production intervals. %
For example, for event $e_1$, produced by segment $\mathit{s5}$ (task $\tau_3$) on core $c_2$, we compute:\\  
$\mathit{Iv_{s5,e_1,1} = bounds\{\phi_{s5, e_1} \, \, and\, \, H_{c_2}.x \leq 20\}:  H_{c_2}.x}$ \\
$\mathit{Iv_{s5,e_1,2} = bounds\{\phi_{s5, e_1} \, \, and\, \, 20 \leq H_{c_2}.x \leq 40\}:  H_{c_2}.x}$\\
where $\phi_{s5, e_1}$ is the same formula as in Sect.~\ref{ca}. \\
We obtain accordingly $\mathit{Iv_{s5,e_1,1} = \{[2,4]\}}$ and $\mathit{Iv_{s5,e_1,2} = \{[22,26],[32,38]\}}$, which correspond to white rectangles in Fig.~\ref{fig4}. %
Second, instead of allowing events to be produced at any moment between their earliest and latest production times, we constrain such production to the exact absolute intervals computed above as follows. %
For each production of an event $e$ by segment $\mathit{s\in JS_\tau}$ within the $k^{th}$ period of $\tau$, we will have as many outgoing edges as $\card{\mathit{Iv_{s,e,k}}}$ from location $s\_k$. %
The guards on these edges will restrict the production of $e$ to the time intervals allowed by the elements of $\mathit{Iv_{s,e,k}}$ (\ie the white rectangles). %
For example, in $\mathit{\mathcal{A}(N_{c_2})}$, the exact abstraction of $\mathit{c_2}$, we will have two outgoing edges from $\mathit{s5\_2}$, one guarded with $\mathit{x \geq 22 \wedge x \leq 26}$ and the other with  $\mathit{x \geq 32 \wedge x \leq 38}$, as illustrated in Fig.~\ref{fig5}. %

Now, we simply compose  $\mathit{\mathcal{A}(N_{c_2})}$ (Fig.~\ref{fig5}) with $\mathit{\mathcal{A}(N_{c_1})}$ and the observer in Fig.~\ref{fig3} (right). %
$\mathit{\mathcal{A}(N_{c_1})}$, the exact abstraction of core $c_1$, is identical to its coarse abstraction shown in Fig.~\ref{fig3} (bottom left) since there are no holes within the earliest and latest production times of $e_2$ (Fig.~\ref{fig4}). %
We obtain, using the query $\mathit{sup\{Obs.recv\}:Obs.x}$, the exact bound $\mathit{eB^\mathit{max}_{e_1, e_2} = 18}$  
which corresponds to the value discussed earlier. 

\subsubsection{On The Distinct-Affinity Assumption}\label{distaff} As said in Sect.~\ref{not}, our approach requires tasks in $T_E$ to have distinct affinities ($\card{T_E} = \card{C_E}$). %
Under this assumption, $\mathit{\mathcal{A}(N_c)}$ is a faithful representative of $c$ \wrt the exact absolute intervals within which $\tau$, the only task in $\mathit{T_E\cap \partition_c}$, produces events. %
This is because such intervals are (i) obtained on the original $N_c$, \ie they include all delays due to scheduling decisions on $c$, and (ii) not influenced by tasks having a different affinity than $\tau$ (since  blocking, if any, is integrated in segments WCETs). %
However, if we lift this assumption, the above is no longer true. %
Let us illustrate this with an example. %
Consider a variant of Example 1 (Sect.~\ref{ex1}) breaking the distinct-affinity assumption: now segment $\mathit{s6}$, in task $\tau_4$, also produces event $e_2$ within the relative interval $[bt_{s6}, wt_{s6}]$. %
Therefore, two tasks producing events (\ie $\tau_3$ and $\tau_4$) have the same affinity ($c_2$). %
The intuitive way to build $\mathit{\mathcal{A}(N_{c_2})}$ would be to compose in parallel $\mathit{\mathcal{A}(TA_{\tau_4})}$ with $\mathit{\mathcal{A}(TA_{\tau_3})}$, two abstractions with the exact production intervals for $e_2$ and $e_1$ by $\tau_4$ and $\tau_3$, respectively. %
$\mathit{\mathcal{A}(TA_{\tau_3})}$ is what we already illustrated in Fig.~\ref{fig5}. %
$\mathit{\mathcal{A}(TA_{\tau_4})}$ will consist of a TA that produces $e_2$ between $18$ and $22$ in each hyperperiod $\mathit{hp_{c_2}}$ (see the analysis in Sect.~\ref{rts-upp}). %
Now, this naive $\mathit{\mathcal{A}(N_{c_2})}$ is not an exact abstraction of $N_{c_2}$: it over-approximates the set of behaviors of $N_{c_2}$ \wrt the production of $e_1$ and $e_2$. %
For example, within the first hyperperiod, $\mathit{\mathcal{A}(N_{c_2})}$ contains the behavior ``$e_1$ produced at $38$ and $e_2$ produced at $21$''. %
This behavior does not exist in $N_c$: if $e_2$ is produced at $21$, then $e_1$ is necessarily produced between $23$ and $26$ (because in this case, $\tau_2$ is preempted at the end of $\mathit{s6}$ and $\mathit{s5}$ executes between $21 + 2 = 23$ and $22 + 4 = 26$, Sect.~\ref{rts-upp}). %
As said in Sect.~\ref{intro}, solving the generic problem of computing exact inter-core bounds is particularly hard. %
We pay the price of exactness through the distinct-affinity restriction. % 

\subsubsection{Automatic Generation} 
We now present algorithms to automatically generate exact abstractions, and begin with setting up some notation. %
We use $(i)$ to denote an element at position $i$ in an ordered set, and $.i$ to access the $i^{th}$ element in a couple. %
The function $\mathit{lb}$ (resp. $\mathit{rb}$) returns the left (resp. right) bound of a real closed interval with natural bounds. %
A compact notation is used for imbricated loops. %
For instance, the $\mathit{for}$ loop in Algorithm~\ref{algca} (lines $5-8$) unfolds into three imbricated $\mathit{for}$ loops, from the outermost to the innermost: $\mathit{for\, \, \tau\, \, in \, \, T_E}$, $\mathit{for\, \, k\, \, in \, \, 1\isep \frac{hp_c}{P_\tau}}$ and $\mathit{for\, \, s\, \, in \, \, JS^{E}_\tau}$. %
Finally, the statements between quotes (`` '') are interpreted as is, with variables replaced by their values. %
For the readability of figures, we sometimes omit clauses in guards that are superfluous. %
For example, if a guard on $s\to s'$ generated by an algorithm is $\mathit{x\geq a \wedge x \leq b}$ and $s$ has an invariant ${x \leq b}$, then the second clause of such guard is sometimes dropped in the figures. %

\paragraph{Computing exact intervals.}
We propose in Algorithm~\ref{algca} a generic procedure to compute $\mathit{Iv}$ variables, which will be used to generate the abstractions. %
First, we generate original $N_c$ networks for each $c\in C_E$ using the automatic generator from FHZ, simplified after verifying schedulability (Sect.~\ref{rts}), and with an additional reference clock for the scheduler;  %
then build $\mathit{SS_c}$, the state space of $N_c$ (lines 1-4). %
Afterwards, we compute, for each $\tau\in T_E$ (with distinct affinity $c$), for each period of $\tau$ in the hyperperiod  $\mathit{hp_c}$ and for each segment $s$ in $\mathit{JS^{E}_\tau}$, the exact intervals of the \emph{first event} produced by $s$, stored in the variables ${\mathit{Iv_{s,s_E(1).1,k}}}$  (lines 5-8). %
This is done through querying the model checker (line 8), on the state space $\mathit{SS_c}$, already computed in line $4$, with $\mathit{bounds}$ properties built using the values of $c$, $\tau$, $k$, $s$ and $s_E$ (line 7). %
For Example 1 (Sect.~\ref{ex1}), we obtain the values of $\mathit{Iv}$ variables discussed in Sect.~\ref{uq}. %

\begin{algorithm}[!ht]
%\setlength{\columnsep}{0.7cm}
%\begin{multicols}{2}
\SetAlCapNameFnt{\scriptsize}
\SetAlCapFnt{\scriptsize}
\scriptsize
%\SetKwInOut{Input}{input}\SetKwInOut{Output}{output}
\SetKwFor{Gen}{compute bounds:}{}{}
\SetInd{0em}{1em}
%\SetKwFor{IntV}{helper variables:}{}{}
%\SetKwFor{Comp}{compute bounds:}{}{}
\SetKwFor{Comp}{generate exact abstractions:}{}{}
%\SetKwRepeat{Repeat}{repeat}{until}
\SetKwComment{tcp}{$\lhd$\mbox{ }}{}
%\SetAlFnt{\tiny}
\SetCommentSty{}
\DontPrintSemicolon

		\For{$c$ in $C_E$}{generate $\mathit{N_c = H_c ||_{\tau \in \partition_c} TA_\tau}$ \tcp*[f]{generate original $N_c$ network (FHZ)}
		
		add clock $x$ to $H_c$
		
		$\mathit{SS_c}$ $\leftarrow$ buildSS($\mathit{N_c}$) \tcp*[f]{call the model checker}
		}

	        \For{$\langle{\tau},{k},{s}\rangle$ in $T_E \times 1\isep \frac{hp_c}{P_\tau} \times \mathit{JS^{E}_\tau}$}{
	        
	        				$c\leftarrow \affinity_\tau$
	        
	        				$\theta$ $\leftarrow$ ``$\mathit{TA_{{\tau}}.{s}\, \, and\, \, lb(s_E(1).2) \leq TA_{{\tau}}.y \leq {rb(s_E(1).2)}\, \, and \, \, }$\newline$\mathit{{(k-1)\cdot P_\tau} \leq H_{{c}}.x \leq {k\cdot P_\tau}}$''
					
	        				$\mathit{Iv_{s,s_E(1).1,k} \leftarrow}$ compute($SS_c$, $\mathit{``bounds\{\theta\}:H_{{c}}.x}$'')  \tcp*[f]{call the model checker}
						         	 
					}

\caption{Computing exact intervals of event production}
\label{algca}
\end{algorithm}

\paragraph{Generating exact abstractions.} %
The algorithm to generate exact abstractions % 
is presented in a gradual manner. %
First, we provide an algorithm under the assumption of \emph{single-job tasks}: each task producing an event has only one job, \ie  $\mathit{\forall \tau \in T_{E} : \card{J_\tau} = 1}$ (and therefore $\mathit{\card{JS^{E}_\tau} = 1}$). 
The procedure under the above assumption is given in Algorithm~\ref{alglocmev}. \\%
\textbf{Single-event segments:}
Let us begin by considering that each segment in $\mathit{\mathcal{JS}_E}$ produces only one event (which is the case of Example 1, Sect.~\ref{ex1}); we ignore therefore for now grayed lines in Algorithm~\ref{alglocmev}. %
 To construct $\mathit{\mathcal{A}(N_c)}$ for some core $c$, the exact abstraction of the only task in $T_E$ allocated to $c$, say $\tau$, we need to generate its clocks, locations and edges. %
The clock is $x$ (line 5). %
Since $\tau$ is single job, we store the only segment in $\mathit{JS^{E}_\tau}$ that produces an event, say $e$, in variable $\mathit{s}$ (line 4). %
Locations $s\_k$ and their invariants are generated for each period of $\tau$ in the hyperperiod $\mathit{hp_c}$ (lines 9-10). %
The constraint over $x$ is driven by the right bound of the last interval in the corresponding $\mathit{Iv}$ variable (line 10). %
Location $\mathit{wait}$, with invariant $\mathit{x \leq hp_c}$, is added (line 13). %
The initial location is $\mathit{s\_1}$ (line 14). %
Afterwards, edges are generated (lines 16-26, 37). %
For each $k^{th}$ period of $\tau$ (up to the hyperperiod) and $i^{th}$ interval (among the possible intervals within which $s$ may produce $e$), the statements in lines 22 and 24 determine the successor of $s\_k$ depending on whether there are more events within the  hyperperiod. %
Then, the auxiliary function $\mathit{edge1()}$ (Algorithm~\ref{algauxfun}) creates an edge accordingly. %
Each $i^{th}$ outgoing edge of location $s\_k$, producing event $e$ in the $\mathit{k^{th}}$ period of $\tau$, has:
\begin{itemize}
\item{} a guard with clock $x$ larger than the left bound of the $i^{th}$ interval in $\mathit{Iv_{s,e,k}}$ and less than its right bound, % 
\item{} an emission (!) on broadcast channel $e$, %
\item{} an empty set of clocks to reset.
\end{itemize}

Line 37 generates an edge from location $\mathit{wait}$ to the event-producing location in the first period, with the guard $\mathit{x = hp_c}$ and a reset over clock $x$. %
For Example 1, Algorithm~\ref{alglocmev} will give accordingly $\mathit{\mathcal{A}(N_{c_2})}$ that we obtained in Fig.~\ref{fig5}, and $\mathit{\mathcal{A}(N_{c_1})}$, illustrated in Fig.~\ref{fig3}, bottom left (we recall that $\mathit{\mathcal{A}(N_{c_1})}$ is identical to $\mathit{\mathcal{CA}(N_{c_1})}$ because of the absence of holes \wrt the production of $e_2$). \\%
\textbf{Multiple-event segments:} 
Now, segments may generate multiple events. %
This part requires a lot of care not to lose the exactness of the abstraction, as we will explain in detail. %

We reason as follows. %
If $s$, the only segment in $\mathit{JS^{E}_\tau}$, produces more than one event ($\card{s_E} > 1$), we need the exact intervals corresponding to the production of the $\mathit{first}$ event in each $k^{th}$ period in the hyperperiod, \ie $\mathit{Iv_{s,s_E(1).1,k}}$, already obtained from the generic Algorithm~\ref{algca}. %
The absolute intervals for producing subsequent events, within the same period, in each \emph{branch} corresponding to each interval in $\mathit{Iv_{s,s_E(1).1,k}}$,  are deduced accordingly. %

\begin{algorithm}[!ht]
%\setlength{\columnsep}{0.7cm}
%\begin{multicols}{2}
\SetAlCapNameFnt{\scriptsize}
\SetAlCapFnt{\scriptsize}
\scriptsize
%\SetKwInOut{Input}{input}\SetKwInOut{Output}{output}
\SetKwFor{Gen}{compute exact intervals:}{}{}
\SetInd{0em}{1em}
%\SetKwFor{IntV}{helper variables:}{}{}
%\SetKwFor{Comp}{compute bounds:}{}{}
\SetKwFor{Comp}{generate exact abstractions:}{}{}
\SetKwFor{Loc}{generate locations:}{}{}
\SetKwFor{Edg}{generate edges:}{}{}
%\SetKwRepeat{Repeat}{repeat}{until}
\SetKwComment{tcp}{$\lhd$\mbox{ }}{}
\SetKwComment{tcpl}{$\lhd$\mbox{ }}{\mbox{ }$\rhd$}
%\SetAlFnt{\tiny}
\SetCommentSty{}
\DontPrintSemicolon
\SetKwFunction{KwEdgeOne}{edge1}
\SetKwFunction{KwEdgeTwo}{edge2}
\SetKwIF{gIf}{gElseIf}{gElse}{\textcolor{gray}{if}}{\textcolor{gray}{then}}{\textcolor{gray}{else if}}{\textcolor{gray}{else}}{\textcolor{gray}{end if}}%

	Algorithm~\ref{algca} \tcp*[f]{Compute exact intervals}

	\tcpl{Compute abstractions}
	
	create exact abstraction $\mathcal{A}(N_E)$ with broadcast channel ``$e$'' for each $e\in E$
	
	\For{${\tau}\in T_E$}{
	$c \leftarrow \affinity_\tau$, $s \leftarrow \mathit{{JS^{E}_\tau(1)}}$
	
	create exact abstraction $\mathit{\mathcal{A}(N_{{c}})}$ with clock set $X = \{x\}$
	
	\textcolor{gray}{\If{$\card{s_E} \neq 1$}{
						$X \leftarrow X \cup \{y\}$
					}}
		
		\Loc{}{	
			\For{${k}$ in $1\isep \frac{hp_c}{P_\tau}$}{

			create location ``${s}\_{k}$'' with invariant ``$x \leq {\mathit{rb(Iv_{s,s_E(1).1,k}(\card{Iv_{s,s_E(1).1,k}}))}}$''
			
			\begingroup
        \color{gray}
			
			\For{$\langle i,j\rangle$ in $1\isep \card{Iv_{s,s_E(1).1,k}} \times 2\isep \card{s_E}$}{
			
			create location ``${s}\_{k}\_i\_j$'' with invariant ``$x \leq rb(Iv_{s,s_E(1).1,k}(i)) + rb(s_E(j).2) - rb(s_E(1).2)\, \,  \&\&$\newline$  y \leq rb(s_E(j).2) - lb(s_E(j-1).2)$''
			
			}
			
			\endgroup
			
			}

			create location ``$\mathit{wait}$'' with invariant ``$\mathit{x \leq {hp_c}}$''

			make location ``$\mathit{{s}\_1}$'' initial
			
			}
			
			\Edg{}{

			\For{$\langle {k}, i \rangle$ in $\mathit{1\isep \frac{hp_c}{P_\tau} \times 1\isep \card{Iv_{s,s_E(1).1,k}}}$}{
				$iv \leftarrow  Iv_{s,s_E(1).1,k}(i)$

					\egIf{\color{gray} $\card{s_E} \neq 1$}{
						\color{gray}
						$succ \leftarrow \mathit{{s}\_{k}\_i\_2}$, $reset \leftarrow \{y\}$
					}{
            \eIf{$k\neq  \frac{hp_c}{P_\tau}$}{
							$succ \leftarrow \mathit{{s}\_{(k+1)}}$
						}{
							$succ \leftarrow \mathit{wait}$
						}
						$reset \leftarrow \varnothing$
					}
          \KwEdgeOne{$s$,$k$,$iv$,$succ$,$reset$}

					\begingroup
        \color{gray}

					\For{$j$ in $2\isep \card{s_E}$}{
						\eIf{$j\neq  \card{s_E}$}{
							$succ \leftarrow s\_{k}\_i\_(j+1)$, $reset \leftarrow \{y\}$
						}{
							\eIf{$k\neq  \frac{hp_c}{P_\tau}$}{
								$succ \leftarrow {s}\_{(k+1)}$
							}{
								$succ \leftarrow wait$
							}
							$reset \leftarrow \varnothing$
						}
						\KwEdgeTwo{$s$,$k$,$i$,$j$,$iv$,$succ$,$reset$}

					}
					\endgroup

		}

	create edge ``($\mathit{wait}$, $x == hp_c$, $\epsilon$, $\{x\}$,  $\mathit{s\_1}$)''
	}

add 	$\mathit{\mathcal{A}(N_{{c}})}$ to $\mathit{\mathcal{A}(N_{{E}})}$
}

\caption{exact abstractions (single-job tasks)}
\label{alglocmev}
\end{algorithm}

\begin{figure}[tb!]
\includegraphics[width=0.75\columnwidth]{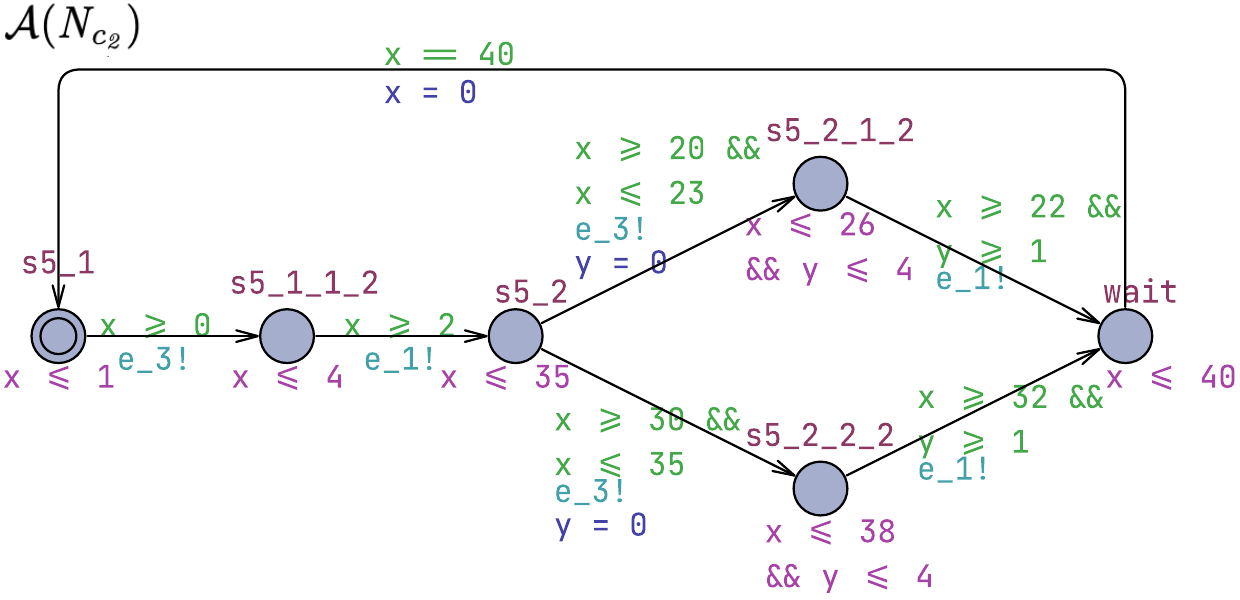}
\centering
\caption{Exact abstraction of core $c_2$ for Example 2 (Sect.~\ref{ex2}), obtained using Algorithm~\ref{alglocmev}.  
The exact abstraction obtained for core $c_1$ is the same as its coarse abstraction in Fig.~\ref{fig3}, bottom left.}    \label{fig7}
\end{figure}

Let us illustrate this through building the abstraction of core $c_2$ from Example 2 (Sect.~\ref{ex2}). %
The exact absolute intervals are computed for the production of $e_3$ by $\mathit{s5}$. %
We obtain accordingly $\mathit{Iv_{s5,e_3,1} = \{[0,1]\}}$ and $\mathit{Iv_{s5,e_3,2} = \{[20,23],[30,35]\}}$. %
For the subsequent events, Algorithm~\ref{alglocmev} creates locations $s\_k\_i\_j$ corresponding to ``producing the $j^{th}$ event in $s_E$, on the $i^{th}$ branch, in the $k^{th}$ period'' (lines 11-12). %
For example, location $\mathit{s5\_1\_1\_2}$ corresponds to producing the second event ($e_1$) on the first branch (here there is only one branch since $\mathit{\card{Iv_{s5,e_3,1}} = 1}$) in the first period. %
Using Algorithm~\ref{alglocmev}, we obtain $\mathit{\mathcal{A}(N_{c_2})}$, illustrated in Fig.~\ref{fig7} ($\mathit{\mathcal{A}(N_{c_1})}$ remains the same as in Fig.~\ref{fig3}, bottom left). %
Obtaining the exact absolute interval of producing an event by $s\_k\_i\_j$ is the hard part. %

\begin{algorithm}[!ht]
  %\setlength{\columnsep}{0.7cm}
  %\begin{multicols}{2}
  \SetAlCapNameFnt{\scriptsize}
  \SetAlCapFnt{\scriptsize}
  \scriptsize
  %\SetKwInOut{Input}{input}\SetKwInOut{Output}{output}
  \SetKwFor{Gen}{compute bounds:}{}{}
  \SetInd{0em}{1em}
  %\SetKwFor{IntV}{helper variables:}{}{}
  %\SetKwFor{Comp}{compute bounds:}{}{}
  \SetKwFor{Comp}{generate exact abstractions:}{}{}
  %\SetKwRepeat{Repeat}{repeat}{until}
  \SetKwComment{tcp}{$\lhd$\mbox{ }}{}
  %\SetAlFnt{\tiny}
  \SetCommentSty{}
  \DontPrintSemicolon

  \SetKwProg{Fn}{Function}{ is}{end}

  \SetKwFunction{KwEdgeOne}{edge1}
  \Fn{\KwEdgeOne{$s$,$k$,$iv$,$s'$,$r$}}{
    create edge ``($\mathit{s\_{k}}$, $\mathit{x \geq  lb(iv)\, \,  \&\& \, \,  x \leq  rb(iv)}$, ${s_E(1).1}!$, $r$,  $s'$)''
  }

  \SetKwFunction{KwEdgeTwo}{edge2}
  \Fn{\KwEdgeTwo{$s$,$k$,$i$,$j$,$iv$,$s'$,$r$}}{
    create edge ``($\mathit{s\_k\_i\_j}, x \geq lb(iv) + lb(s_E(j).2) -  lb(s_E(1).2) \, \,  \&\&$\newline$y \geq lb(s_E(j).2)-rb(s_E(j-1).2), s_E(j).1!, r,  s'$)''
  }

  \caption{Auxiliary functions}
  \label{algauxfun}
  \end{algorithm}

First, let us ignore constraints over clock $y$ in Fig.~\ref{fig7}. %
Let $s$ be a segment and %
$\mathit{s_E = \{(e,[a,b]), (e',[a',b'])\}}$. % 
We know that, in any execution scenario, if $e$ is produced by $s$ in the absolute interval $[f,h]$, $e'$ will follow within $[f+c,h+d]$ where $c = a'-a$ and $d = b'-b$. %
This is guaranteed because segments may not be preempted (Sect.~\ref{rts}) and  
$[c,d]$ is necessarily non-empty (Sect.~\ref{not}). %
Constraints over $x$ will therefore model $[f+c,h+d]$, where each $[f,h]$ interval within the $k^{th}$ period is an interval in $\mathit{Iv_{s,e,k}}$, which corresponds to a \emph{branch}, as introduced above. %
In Fig.~\ref{fig7}, the invariant constraint of $\mathit{s5\_2\_1\_2}$ over $x$, for instance, is $x \leq 26$, where, using the notation above, $h+d = 26$, with $h = 23$ (the right bound of the first element in $\mathit{Iv_{s5,e_3,2}}$) and $d = b'-b = 4 - 1 = 3$, $4$ (resp. $1$) being the right bound of $\mathit{s5_E(2).2}$ (resp. $\mathit{s5_E(1).2}$). %
The guard $x \geq 22$ on the outgoing edge of $\mathit{s5\_2\_1\_2}$ is obtained similarly. %
We get the same constraints over $x$ for the production of $e_1$ as in Fig.~\ref{fig5}, which is expected since the producer and relative production intervals of $e_1$ are the same  as in Example 1 (Sect.~\ref{ex1}). %

However, these constraints are not enough to guarantee an exact abstraction. %
In our example, following only the constraints on $x$ in Fig.~\ref{fig7},  $\mathit{s5\_2}$ (resp. $\mathit{s5\_2\_1\_2}$) can produce $e_3$ (resp. $e_1$) at any instant between $20$ and $23$ (resp. $22$ and $26$). %
In the original network, however, it is not possible to produce $e_3$ then $e_1$, both at \eg $22$, because their relative production intervals $[0,1]$ and $[2,4]$ are separated by at least one time unit. %
This difficulty arises from the fact that a range within an absolute interval corresponds to a \emph{sliding} relative interval: values in $[20,22]$ may all correspond to the relative $0$ for $e_3$. %
To remedy this, we conjunct constraints on $x$ with constraints on clock $y$ (created in line 7). %
The latter is reset when an event $e$ is produced, and constrains the production of a  subsequent event $e'$ using information from their relative production intervals. %
For example, $\mathit{s5\_2\_1\_2}$ may produce $e_1$ within its absolute interval iff it obeys the minimum delay, \ie $2 - 1 = 1$, and the maximum delay, \ie $4-0 = 4$ separating $e_1$ and $e_3$. %
The conjunction of constraints works nicely, preventing $\mathit{s5\_2\_1\_2}$ to produce $e_1$ too early or too late. %
On the one hand, constraints over $x$ forbid $e_1$ to occur $1$ (resp. $4$) time units after $e_3$ if the production of $e_3$ corresponds exclusively to the relative instant $0$ (resp. $1$). %
That is, if $e_3$ occurs at $20$ (resp. $23$), which corresponds necessarily to the relative instant $0$ (resp. $1$), the guard $x\geq 22$ (resp. the invariant $x\leq26$) will prevent $e_1$ to be produced at $21$, \ie too early (resp. $27$, \ie too late). %
On the other hand, constraints on $y$ will prevent $\mathit{s5\_2\_1\_2}$ to produce $e_1$ \eg at $26$ if $e_3$ is produced at $21$ (too late) or at $22$ when $e_3$ is produced at the same time (too early). %
Note that, for the readability of Fig.~\ref{fig7}, constraints on $y$ are removed when superfluous. %

\begin{figure}[tb!]%
\centering
\includegraphics[width=0.75\columnwidth]{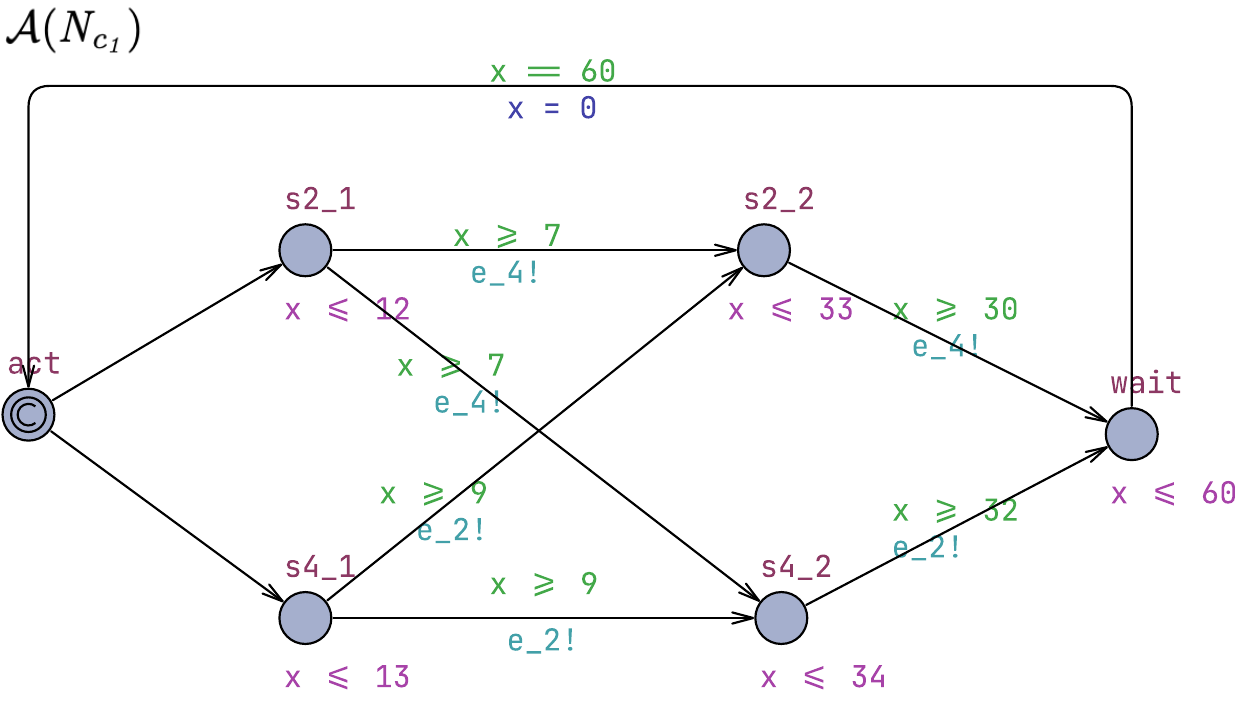}
\centering
\caption{Exact abstraction of core $c_1$ for Example 3 (Sect.~\ref{ex3}), obtained using Algorithm~\ref{algfin}.  
The exact abstraction obtained for core $c_2$ is the same as in Fig.~\ref{fig7}.}    \label{fig8}%
\end{figure}

\textbf{General case:}
Finally, we lift the single-job assumption in Algorithm~\ref{algfin}. %
The main difference is when task $\tau$ is multiple job (lines 6-28). %
The difficulty here is that multiple segments (in different jobs) may produce events, but we may only have one initial location in a TA. %
For this, we use an initial, committed location $\mathit{act}$ (line 14) that allows to atomically transit to one of the event producers in the first period, through edges $\mathit{act\to s\_1}$, where $s$ is an element of $\mathit{JS^{E}_\tau}$ (line 20). 
For the remaining edges, it suffices to iterate over $\mathit{JS^{E}_\tau}$ and reuse the corresponding instructions from Algorithm~\ref{alglocmev} (line 21). %
Then, we need to consider paths corresponding to producing an event within one job in one period, and producing an event within a different job in the next period (lines 22-27). %
For Example 3 (Sect.~\ref{ex3}), Algorithm~\ref{algfin} generates $\mathit{\mathcal{A}(N_{c_1})}$ illustrated in Fig.~\ref{fig8}. %
For single-job tasks, we take a shortcut reusing lines from Algorithm~\ref{alglocmev}, therefore generating, for Example 3, the same $\mathit{\mathcal{A}(N_{c_2})}$ illustrated in Fig.~\ref{fig7}. %

\begin{algorithm}[!ht]
%\setlength{\columnsep}{0.7cm}
%\begin{multicols}{2}
\SetAlCapNameFnt{\scriptsize}
\SetAlCapFnt{\scriptsize}
\scriptsize
%\SetKwInOut{Input}{input}\SetKwInOut{Output}{output}
\SetKwFor{Gen}{compute exact intervals:}{}{}
\SetInd{0em}{1em}
%\SetKwFor{IntV}{helper variables:}{}{}
%\SetKwFor{Comp}{compute bounds:}{}{}
\SetKwFor{Comp}{generate exact abstractions:}{}{}
\SetKwFor{Loc}{generate locations:}{}{}
\SetKwFor{Edg}{generate edges:}{}{}
%\SetKwRepeat{Repeat}{repeat}{until}
\SetKwComment{tcp}{$\lhd$\mbox{ }}{}
\SetKwComment{tcpl}{$\lhd$\mbox{ }}{\mbox{ }$\rhd$}
%\SetAlFnt{\tiny}
\SetCommentSty{}
\DontPrintSemicolon
\SetKwFunction{KwEdgeOne}{edge1}
\SetKwFunction{KwEdgeTwo}{edge2}

	Algorithm~\ref{algca}  \tcp*[f]{Compute exact intervals}
		
	\tcpl{\quad Compute abstractions}
	
	create exact abstraction $\mathcal{A}(N_E)$ with broadcast channel ``$e$'' for each $e\in E$
	
	\For{${\tau}\in T_E$}{
		
		\eIf{$\card{JS^{E}_\tau} = 1$}{
		
		Algorithm~\ref{alglocmev}, lines 4-37
		
		}{
		
		$c\leftarrow \affinity_\tau$
		 
		Create exact abstraction $\mathit{\mathcal{A}(N_{{c}})}$ with clock set $X \leftarrow\{x\}$
		
		\For{$s\in JS^{E}_\tau$}{
			
			\If{$\card{s_E} > 1$}{
			
				$X \leftarrow X \cup \{y\}$
				
				break
			}
		}
		
		\Loc{}{	

			create committed initial location ``$act$''
			
			create location $\mathit{wait}$ with invariant ``$x \leq hp_c$''
			
			\For{$s$ in $JS^{E}_\tau$}{
			
			Algorithm~\ref{alglocmev}, lines 9-12
			
			}
			}
			
			\Edg{}{
			\For{$s$ in $\mathit{JS^{E}_\tau}$}{
			
			create edge ``($\mathit{act}$, $\top$, $\epsilon$, $\varnothing$,  $\mathit{s\_1}$)''
			
			Algorithm~\ref{alglocmev}, lines 16-36
						
			}
			
			\For{$\langle s, k, iv, s'\neq s\rangle$ in $\mathit{JS^{E}_\tau \times 1\isep \frac{hp_c}{P_\tau} - 1 \times Iv_{s,s_E(1).1,k} \times JS^{E}_\tau}$}{
			
			\eIf{$\card{s_E} = 1$}{
			
			\KwEdgeOne{$s$,$k$,$iv$,$\mathit{{s'}\_{(k+1), \varnothing}}$}
			}{
			
			$i \leftarrow index(iv)$
			
			\KwEdgeTwo{$s$,$k$,$i$,$\card{s_E}$,$iv$,$\mathit{{s'}\_{(k+1), \varnothing}}$}
			}
			}
			
			create edge ``($\mathit{wait}$, $x == hp_c$, $\epsilon$, $\{x\}$,  $\mathit{act}$)''
			}

}			
add 	$\mathit{\mathcal{A}(N_{{c}})}$ to $\mathit{\mathcal{A}(N_{{E}})}$
}
			
\caption{exact abstractions (general case)}
\label{algfin}
\end{algorithm}

Note that, in the particular case of multiple-job tasks producing events, the validity of requirements is very important. %
Consider \eg Example 3 (Sect.~\ref{ex3}) with the requirement $\mathit{Req_1}$ ``compute the maximal bound between each production of $e_4$ and the next production of $e_3$''. %
$\mathit{Req_1}$ is valid, because it seeks the maximum amount of time between any production of $e_4$ (given that it takes place) and the subsequent $e_3$ (which always happens). %
In contrast, $Req_2$, corresponding to $Req_1$ where $e_3$ and $e_4$ are switched, is invalid. %
Indeed, such bound does not exist: there are infinite executions where $e_4$ does not occur at all, \ie following the path where $\tau_2$ never executes the job $\mathit{\{s2,s3\}}$. %
This is not a verification issue, but a design issue, and does not have an influence on the abstraction's exactness. %
The trickier, and more dangerous case, is when the user has an invalid requirement in mind, and tries to build an exact abstraction for a multiple-job task where some jobs produce no events at all. %
Our implementation of Algorithm~\ref{algfin} anticipates this case (details in Sect.~\ref{impl}). 

\section{Evaluation}\label{eval}
\subsection{Technicalities}\label{impl}

We automate our approach through a full implementation of the algorithms presented in Sect.~\ref{appr}. %
The code is written in OCaml and is publicly available. %
Several technicalities are addressed. 

\paragraph{Improving scalability.} Algorithm~\ref{algca} contains two potential sources of costly computation: (i) the heavy use of clock constraints in the $\mathit{bounds}$ query (line 3) and (ii) the multiple calls to the model checker (lines 7-8). %
Point (i) stems from the fact that in $\mathit{sup}$ and $\mathit{inf}$, and their $\mathit{bounds}$ generalization, it is recommended to use as few clock constraints as possible (in the enclosed state formula) for a better scalability. %
In our implementation of Algorithm~\ref{algca}, we drastically minimize the number of clock constraints in $\mathit{bounds}$ queries as well as the number of invocations of the model checker using the following trick. % 
Instead of adding a clock to the scheduler $H_c$ to get an absolute reference of time, we compose $N_c$ with a TA $\mathit{Ref}$ with one clock $x$ and one location $\mathit{hper}$ with the invariant $x\leq hp_c$. %
This creates an artificial \emph{timelock} at exactly $hp_c$, therefore allowing to stop the exploration when the reference clock hits the hyperperiod limit. %
Then, instead of querying the model checker $k$ times for each event-producer within each task, we simply replace clock constraints over $x$ in $H_c$ with the proposition $\mathit{Ref.hper}$ and compute only one time the bounds on $\mathit{Ref.x}$ per event-producing segment. %
We get therefore all the absolute production intervals of the first event in each segment within the hyperperiod at once, then simply group them according to the period to which they belong. %
For Example 1 (Sect.~\ref{ex1}), for instance, in order to build the exact abstraction $\mathit{\mathcal{A}(N_{c_2})}$, we query only once: $$\mathit{Iv = bounds\{\phi_{s5, e_1}\, \, and\, \, Ref.hper\}: Ref.x}$$
To obtain $Iv = \{[2,4], [22,26],[32,38]\}$, then group the intervals per period of $\tau_3$ to get $\mathit{Iv_{s5,e_1,1} = \{[2,4]\}}$ and $\mathit{Iv_{s5,e_1,2} = \{[22,26],[32,38]\}}$. %

\paragraph{Guaranteeing exactness under invalid requirements.} In our implementation of Algorithm~\ref{algfin}, in the case of multiple-job tasks producing events, we always issue a warning, asking the user to make sure their requirement is valid (\ie they are not trying to compute a bound that does not exist). %
Moreover, we check that, in such multiple-job tasks, each job has a segment producing an event. %
If this is not the case, the procedure is aborted. %
To explain the reason behind this, let us consider a variation of Example 3 (Sect.~\ref{ex3}) where $\tau_2$ produces no longer $e_2$ and the requirement $Req$ ``compute the maximal bound between each production of $e_3$ and the next production of $e_4$''. %
This requirement is clearly invalid, for the same reasons depicted at the end of Sect.~\ref{appr}. %
The problem here is, if the user chooses to generate the abstraction anyway, it will be no longer exact given the requirement $Req$: the path corresponding to the job $\{s2,s3\}$ will no longer exist and the verification would give a value for the non-existent bound defined by $Req$. %
Therefore, we generate no abstraction in this case and ask the user to provide an event producer for every job. % 
This choice is deliberately conservative, as we seek to guarantee the exactness of the abstraction regardless of whether the requirement is valid or not. %
It is still possible to ignore this safeguard and generate the abstraction anyway (see below). %

\paragraph{Options.} The \textsf{-xta} option allows to use existing TA files for $N_c$ networks (\ie skipping line 2 in Algorithm~\ref{algca}), and the \textsf{-verbose} option to visualize the computation of intervals and their grouping within periods. %
The \textsf{-force} option allows to generate the abstraction 
if at least one multiple-job task $\tau\in T_E$ fails the test above (at least one of its jobs contains no event-producing segment). %
In this case, it is up to the user to make sure that their requirement guarantees the exactness of the generated abstraction. % 

\subsection{Experiments}
To demonstrate the scalability of our approach, we compute exact inter-core bounds of various types on the WATERS 2017 industrial challenge~\cite{hamann2017waters}. %
The reported results have been obtained using \uppaal 5.1.0 (with the state-space representation option set to DBM) on a mid-range computer  with an Intel Core i9 processor, 10 cores and 32 GiB of RAM. %
The results are reproducible in a fully automatic manner\footnote{\url{https://gitlab.math.univ-paris-diderot.fr/mzhang/exact-abstraction}}. %

The challenge underlies a real automotive quad-core RTS. %
Our approach is applicable to the subsystem with cores $c_1$ and $c_2$, all tasks allocated to which are periodic. %
Model checking this subsystem is particularly challenging. %
First, the number of interleavings is extremely high: nine tasks, with $897$ segments overall (Table~\ref{tab:waters}). %
Second, WCETs and BCETs of segments (not reported in Table~\ref{tab:waters}) have a nanosecond resolution, and 
the difference between the smallest and the largest timing constraint is about $1$ billion time units. %
Tasks segments use data \emph{labels} to communicate. %
$N_{c_1}$ and $N_{c_2}$ are both schedulable given a frequency of $\mathit{400\, MHz}$. %
FHZ verified $N_{c_2}$ apart, computing precise WCRTs of tasks in $\partition_{c_2}$ with blocking due to labels sharing taken into account. %
All tasks in the challenge are single job. %

We want to compute various minimum and maximum exact bounds between writing a label $\mathit{lbl_x}$, by a segment $s$ in a task allocated to one core, and writing another label $\mathit{lbl_y}$ by a segment $s'$, in a task allocated to the other core, such that $s'$ writes $\mathit{lbl_y}$ based on a prior reading of $\mathit{lbl_x}$. %
We discuss the results for labels $x = 1327$ and $y = 4164$, where segment $\mathit{Runnable\_6660\_us\_1}$ in task $\mathit{Angle\_Sync}$ (core $c_1$) writes $\mathit{lbl_{1327}}$, and segment $\mathit{Runnable\_50\_ms\_2}$ in task $\mathit{T\_50}$ (core $c_2$) reads $\mathit{lbl_{1327}}$ and writes $\mathit{lbl_{4164}}$ accordingly. %
For simplicity, we use $1$ instead of $1327$ (for $x$), $2$ instead of $4164$ (for $y$), $s1$ instead of $\mathit{Runnable\_6660\_us\_1}$ and $s2$ instead of $\mathit{Runnable\_50\_ms\_2}$. %
Using \emph{implicit communication semantics}, a segment reads (resp. writes) labels at the beginning (resp. end) of its execution. %
Accordingly, $\mathit{E = \{w_1, r_1, w_2\}}$, $\mathit{C_E = \{c_1, c_2\}}$, $\mathit{T_E = \{Angle\_Sync, T\_50\}}$, \\ $\mathit{JS^{E}_{Angle\_Sync} = \{s1\}}$, $\mathit{JS^{E}_{T\_50} = \{s2\}}$, $\mathit{s1_E = \{(w_1, [bt_{s1}, wt_{s1}])\}}$ and \\ $\mathit{s2_E = \{(r_1, [0,0]),(w_2,[bt_{s2}, wt_{s2}])\}}$, and the requirement is ``compute exact bounds between each $w_1$ and the next $w_2$ with $r_1$ happening in between''. %
Note the relative interval associated with $r_1$, equal to $[0,0]$. %
Since $r_1$ occurs before $w_2$ within the same segment, and given the requirement above, the latest relative instant corresponding to reading $\mathit{lbl_1}$ is unimportant: what matters is that $r_1$ actually happens between $w_1$ and $w_2$. %

\begin{table}[tb]
\scriptsize
\center%
\begin{tabular}{|l|r|l|l|l|}%
\hline
Task 
               & \thead{Period \\ ($\mathit{\times 10^{-6} s}$)}
                       & Priority
                                 & \# Segments
                                         & Affinity \\
\hline
T\_2    &     2\, 000 & 6       &  28   & $c_2$\\
T\_5    &     5\, 000 & 5       &  23   & $c_2$\\
T\_20   &    20\, 000 &  4       & 307   & $c_2$\\
T\_50   &    50\, 000 &  3       &  46   & $c_2$\\
T\_100  &   100\, 000 &  2       & 247   & $c_2$\\
T\_200  &   200\, 000 &  1       &  15   & $c_2$\\
T\_1000 & 1\, 000\, 000 &  0       &  44   & $c_2$\\
\hline
\task{Angle\_Sync} & 6\, 660 & 1 & 146 & $c_1$\\
\task{T\_1} & 1\, 000 & 0 & 41 & $c_1$\\ 
\hline
\end{tabular}
\vspace{0.2cm}
\caption{WATERS 2017 industrial challenge information (core $c_1$ and $c_2$).}
\label{tab:waters}
\end{table}

Leveraging the power of observers, we answer 
the requirement above, following different latency semantics~\cite{feiertag2009compositional}. %
The \emph{first-to-first (FF)} semantics corresponds simply to ``between each production of $w_1$ and the next $w_2$ with $r_1$ happening in between'' (this semantics is what we used in Sect.~\ref{ca}, with two events, to illustrate the hole phenomenon). %
The \emph{last-to-first (LF)} semantics exclude overwritten $w_1$ events. %
That is, FF excluding executions where at least one $w_1$ occurs between $w_1$ and $r_1$. %
The exact minimum and maximum FF and LF bounds are denoted, respectively, $\mathit{eBF^{min}_{w_1,r_1,w_2}}$ and $\mathit{eBF^{max}_{w_1,r_1,w_2}}$, $\mathit{eBL^{min}_{w_1,r_1,w_2}}$ and $\mathit{eBL^{max}_{w_1,r_1,w_2}}$. %

\begin{figure}[tb!]
\centering
\includegraphics[width=0.58\columnwidth]{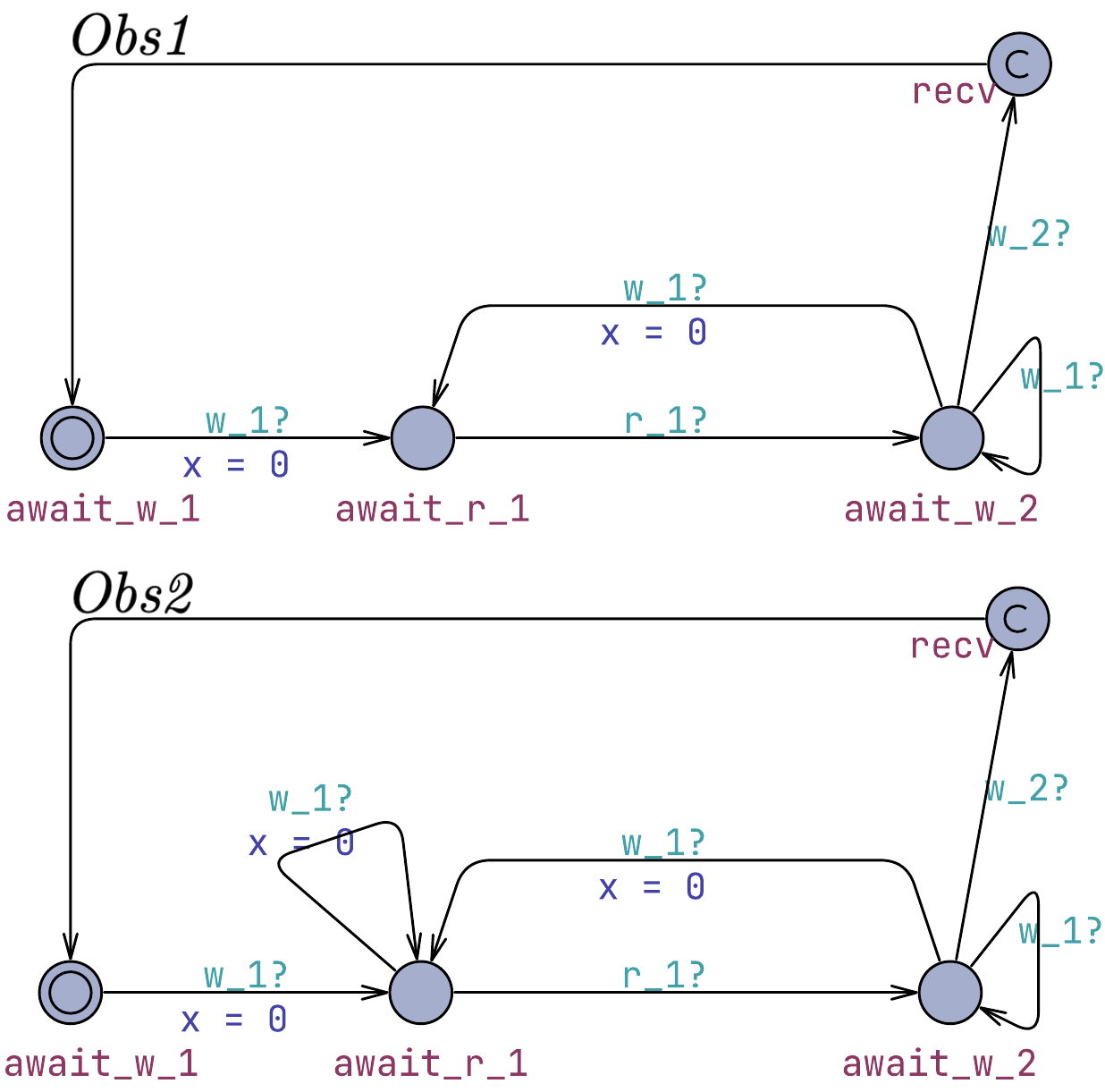}
\centering
\caption{Observers for FF and LF bounds.}    \label{fig9}
\end{figure}

To compute FF bounds, we use $\mathit{Obs1}$ (Fig.~\ref{fig9}, top). %
It awaits for a chain of events $w_1$ (edge $\mathit{await\_w\_1\to await\_r\_1}$, with a reset over clock $x$) followed by $r_1$ (edge $\mathit{await\_r\_1\to await\_w\_2}$) followed by $w_2$ (edge $\mathit{await\_w\_2\to recv}$). %
Events $w_1$ that may occur while awaiting $r_1$ are safely ignored. %
In contrast, events $w_1$ that may occur while awaiting $w_2$ must be considered. %
The non determinism over the edges $\mathit{await\_w\_2\to}$ synchronized on $w_1$ allows to either take such $w_1$ events into account  or ignore them. %
Therefore, all the chains $w_1\rightarrowtail r_1\rightarrowtail w_2$, such that  further occurrences of $w_1$ between $w_1$ and $r_1$ are ignored, are taken into account (FF semantics). %
For LF bounds, $\mathit{Obs2}$ (Fig.~\ref{fig9}, bottom) is similar to $\mathit{Obs1}$, with the difference that it takes into account overwriting $w_1$ through resetting clock $x$ whenever an event $w_1$ occurs while waiting for $r_1$. %
Therefore, to get exact FF and LF bounds, \ie respectively, $\mathit{eBF^{min}_{w_1,r_1,w_2}}$, $\mathit{eBF^{max}_{w_1,r_1,w_2}}$, and $\mathit{eBL^{min}_{w_1,r_1,w_2}}$ and $\mathit{eBL^{max}_{w_1,r_1,w_2}}$, it suffices to query $\mathit{inf\{Obs1.recv\}:Obs1.x}$, $\mathit{sup\{Obs1.recv\}:Obs1.x}$, and $\mathit{inf\{Obs2.recv\}:Obs2.x}$, $\mathit{sup\{Obs2.recv\}:Obs2.x}$. 

\begin{table}
\parbox{.45\columnwidth}{
\centering
\begin{tabular}{|l|r|r|}
\hline
Abstraction & \thead{Cost \\ (s)} & \thead{Cost \\(GiB)}\\
\hline
$\mathit{\mathcal{A}(N_{c_1})}$ & 109 &  0.56 \\
\hline
$\mathit{\mathcal{A}(N_{c_2})}$ & 122 & 3.3  \\
\hline
\end{tabular}
\vspace{0.2cm}
\caption{Cost (time and memory) of abstractions' computation}
\label{cost}
}
\hfill
\parbox{.45\columnwidth}{
\centering
\begin{tabular}{|l|r|}
\hline
Bound&Value ($10^{-9}$ s)\\
\hline
$\mathit{eBF^{min}_{w_1,r_1,w_2}}$ & 1\, 766\, 230 \\
\hline
$\mathit{eBF^{max}_{w_1,r_1,w_2}}$ & 56\, 808\, 228 \\
\hline
$\mathit{eBL^{min}_{w_1,r_1,w_2}}$ & 14\, 080 \\
\hline
$\mathit{eBL^{max}_{w_1,r_1,w_2}}$ & 7\, 074\, 911 \\
\hline
\end{tabular}
\vspace{0.2cm}
\caption{Exact FF and LF bounds}
\label{res}
}
\hfill
\end{table}

We first apply the direct method from FHZ, \ie try to compute the bounds on the parallel composition $\mathit{N_E || Obs}$ with $N_E = N_{c_1} || N_{c_2}$ and $\mathit{Obs\in \{Obs1, Obs2\}}$. %
The verification fails with all 32 GiB of RAM exhausted in less than 15  minutes. %
Then, we use our approach to generate $\mathcal{A}(N_E) = \mathcal{A}(N_{c_1}) || \mathcal{A}(N_{c_2})$. %
The cost of computing $\mathcal{A}(N_{c_1})$ and $\mathcal{A}(N_{c_2})$, mainly driven by model checking $N_{c_1}$ and $N_{c_2}$ apart (to compute the exact intervals), is reported in Table~\ref{cost}:  %
 less than four minutes and less than 4 GiB of RAM overall. %
We compose $\mathcal{A}(N_E)$ with each observer and compute FF and LF bounds (Table~\ref{res}); the cost of such computation is negligible (less than 1 second, and less than 30 MB of RAM). %

\subsection{Discussion} We efficiently compute, in a fully automatic manner, various exact (to the nanosecond) inter-core bounds on the real  WATERS 2017 industrial challenge. %
To the best of our knowledge, this is the first work that does so. %
These exact bounds significantly reduce pessimism. %
For \eg $\mathit{eBF^{max}_{w_1,r_1,w_2}}$, its value using LET semantics, a popular choice in automotive industry (Sect.~\ref{rw}), is $96\, 920\, 000$, \ie $> 40$ $\mathit{ms}$ larger than the one we computed (Table~\ref{res}). % 
Yet, our technique is restricted to a scheduling policy, and by the distinct-affinity assumption. %
LET-based approaches are therefore more generic, but more pessimistic than ours (Sect.~\ref{rw}). %

\section{Related Work}\label{rw}

Due to the state-space explosion problem, the literature related to inter-core bounds is dominated by analytical approaches, %
 mostly specific to end-to-end latencies. %
The latter typically refer to maximum bounds in cause-effect chains (Becker et al.~\cite{becker2017end}), separating sensor reading and actuation. %
Feiertag et al.~\cite{feiertag2009compositional} propose in a seminal paper a number of algorithms for end-to-end latencies on single-core platforms, for which they define precise semantics. %
The pessimism of upper bounds in cause-effect chains is reduced in Kloda et al.~\cite{kloda2018latency} and Girault et al.~\cite{girault2018improving}, under specific workload and scheduling assumptions (P-FP preemptive scheduling for the former, and event-triggered workloads on single-core platforms for the latter). %
Martinez et al.~\cite{martinez2020end} and Günzel et al.~\cite{gunzel2023compositional} provide upper bounds for end-to-end latencies with different semantics under different communication models for automotive systems. %
 Due to the dependency of existing algorithms (to over-approximate end-to-end latencies) on scheduling assumptions, communication models based on the Logical Execution Time (LET)~\cite{henzinger2001embedded} gained significant popularity ~\cite{martinez2018analytical}. %
Using \emph{publishing points} of read/write events~\cite{martinez2018analytical,martinez2020end}, they are oblivious to scheduling assumptions. Therefore, the computed bounds remain valid regardless of tasks execution order. %
The price to pay is however larger bounds, as we have exemplified in Sect.~\ref{eval}. %

Our approach shares a central assumption with the related work above: they all consider a schedulable RTS. %
Besides, all such works, excluding the ones on LET, assume that a faithful model, that \eg integrates blocking, exists. %
However, we compute exact bounds, whereas the works above over-approximate them. %
Also, our model is not restricted to cause-effect chains. %
In particular, in a cause-effect chain, a task consumes one event and produces another one, which is consumed by the next task and so on. %
Our model does not make such assumption, and is therefore suitable for bounds where \eg two tasks in parallel produce events that are used by a subsequent task (see \eg the drone application in~\cite{{DBLP:journals/firai/FoughaliZ22}}). %
Moreover, we preserve the generality of model checking, therefore allowing for extensions to verify other properties. %
On the downside, the distinct-affinity assumption is a strong one. % 
Further, compared to LET-based approaches, our model is restricted to a specific, even though popular, scheduling model. %
The last two points are the price we pay for exactness. %

\section{Conclusion}\label{concl}

We presented in this paper a novel scalable approach to compute exact inter-core bounds in a schedulable RTS. %
Our technique, fully automated, is based on an algorithm that computes exact abstractions through a new query that we implemented in the state-of-the-art model checker \uppaal. %
The scalability of our algorithms is demonstrated on the WATERS 2017 industrial challenge, on which we successfully computed various exact inter-core bounds in less than four minutes and with less than 4 GiB of memory. However, our work is restricted to periodic tasks under P-FP scheduling with limited preemption, and more importantly constrained by the distinct-affinity assumption. %
In future work, we plan to extend our method to classical cause-effect chains, therefore relaxing the distinct-affinity assumption. %
This will allow us to provide engineers with two different solutions based on their needs. %

\section{Acknowledgements} This work has received support under the program ``Investissement d'Avenir'' launched by the French Government and implemented by ANR, with the reference ``ANR‐18‐IdEx‐000'' as part of its program ``Emergence''. %
MAF, the main author, thanks ANR  MAVeriQ (ANR-20-CE25-0012) and  ANR-JST CyPhAI (ANR-20-JSTM-0001/JPMJCR2012).

%\section{Implementation}\label{impl}
%
\bibliographystyle{splncs04}
\bibliography{bib}

%\begin{subappendices}
%\renewcommand{\thesection}{\Alph{section}}
%\input{appendix.tex}
%\end{subappendices}
%
\end{document}